\documentclass[aps,prx,twocolumn,superscriptaddress,showpacs,final]{revtex4-2} 
\usepackage[normalem]{ulem}
\usepackage{amssymb}
\usepackage{suffix}
\usepackage{mathtools}
\usepackage[utf8]{inputenc}
\usepackage{booktabs}
\usepackage{cases}
\usepackage[multiple]{footmisc}
\usepackage{dcolumn}
\usepackage{color,soul}
\usepackage{rotating}
\usepackage{perpage}
\usepackage{xcolor}
\usepackage{soul}
\usepackage{tikz}
\usepackage[T1]{fontenc}
\usepackage{etoolbox}
\usepackage{graphics}
\usepackage{siunitx}
\usepackage{hyperref}
\usepackage{float}
\usepackage{multirow}
\usepackage{mathtools}
\usepackage{bm}
\usepackage{url}

\usepackage{tikz}
\usepackage{tikz-3dplot}
\usepackage [outdir=./]{epstopdf}
\usepackage{bbold}
\usepackage{changes}
\newcommand{\az}{{\alpha_z}}
\allowdisplaybreaks
\begin{document}
\title{Floquet states and optical conductivity of an irradiated two dimensional topological insulator}

\author{S. Sajad Dabiri}
\address{Department of Physics, Shahid Beheshti University, 1983969411 Tehran, Iran}
\author{Hosein Cheraghchi}
\email{cheraghchi@du.ac.ir}
\address{School of Physics, Damghan University, P.O. Box 36716-41167, Damghan, Iran}
\address{School of Physics, Institute for Research in Fundamental Sciences (IPM), 19395-5531, Tehran, Iran}

\author{Ali Sadeghi}
\address{Department of Physics, Shahid Beheshti University, 1983969411 Tehran, Iran}
\address{School of Nano Science, Institute for Research in Fundamental Sciences (IPM), 19395-5531, Tehran, Iran}

\date{\today}
\vspace{1cm}
\newbox\absbox
\begin{abstract}
We study the topology of the Floquet states and time-averaged optical conductivity of the lattice model of a thin topological insulator subject to a circularly polarized light using the extended Kubo formalism. Two driving regimes, the off-resonant and on-resonant, and two models for the occupation of the Floquet states, the ideal and mean-energy occupation, are considered.
In the ideal occupation, the real part of DC optical Hall conductivity is shown to be quantized while it is not quantized for the mean energy distribution. The optical transitions in the Floquet band structure depend strongly on the occupation and also the optical weight which consequently affect all components of optical conductivity. At high frequency regime, we present an analytical calculation of the effective Hamiltonian and also its phase diagram which depends on the tunneling energy between two surfaces. The topology of the system shows rich phases when it is irradiated by a weak on-resonant drive giving rise to emergence of anomalous edge states. 
\end{abstract}
\maketitle

\section{Introduction}
\added{Three-dimensional} topological insulators (TI) are quantum materials having some surface %/edge 
states within the band gap which originate from the topological properties of the bulk band structure.
Similarly, a two-dimensional TI has \emph{edge} states appearing within the band gap. 
These edge states have traces in measurement of some quantities such as quantum Hall conductivity.
At zero probe frequency (dc measurement), the Hall conductance of a TI is quantized as $Ch (e^2/h)$ where $Ch$ is the Chern number i.e. the number of the edge modes evaluated from the %bulk
wave function~\cite{klitzing,thouless}. However, such topological phenomena and the edge states are absent in the {\it ac} quantum Hall effect in which robust step-like structures are nontrivial~\cite{Aoki2009}. 

Chern insulators are classified in the category of the TIs with broken time-reversal symmetry in which the quantized edge conductivity emerges without any use of external magnetic field~\cite{haldane}. For noninteracting fermionic systems, there is a complete classification of topological phases according to their symmetries~\cite{chiu}. The number of materials possessing topological properties is limited, so it is desirable to find a way to engineer such properties in
different materials. A dynamical way to engineer quantum properties of materials %which has 
recently attracted great attention, is applying time-periodic perturbations. %It is shown that 
By irradiating circular polarized light on graphene, it is possible to open %the bulk 
an energy gap and make a state called Floquet Topological insulator (FTI)~\cite{oka2009,kitagawa}. Experimentally, Floquet-Bloch states and also FTIs have been realized in different systems such as optical lattices~\cite{opt}, graphene~\cite{graphene} and the surface of a 3D TI~\cite{gedik}. 

The topology in driven systems represents richer features than in the static ones. 
For instance, we have recently shown that \added{by} variation of the light \deleted{such as polarization, amplitude and frequency} or the system parameters \deleted{such as magnetization, external magnetic field and structure inversion asymmetry}, a rich phase diagram emerges in irradiated thin TIs ~\cite{dabiri1,dabiri2}. In comparison to ordinary TIs, in the FTIs dynamical gaps emerge which can host \emph{anomalous} edge
states~\cite{rudner}. To characterize this type of edge states, beside the Chern number of bands, the winding number is also defined as a new topological invariants stemming from the symmetry of the Hamiltonian in time~\cite{rudner}. A periodic table for FTIs~\cite{roy} and the associated invariants~\cite{FTIinvariants} is already present. % have been already classified. 
The anomalous edge states typically appear at the on-resonant drives when the drive frequency is lower than the bandwidth of the system.

For high frequency drives, the Chern numbers of bands are sufficient to describe the chiral edge states in the gap at zero energy. In this case, at a drive with weak intensity, it is convenient to use an effective static Hamiltonian which gives the stroboscopic evolution of the system. If the photon energy is larger than any characteristic energy of the system e.g. energy change arising from single-scattering processes in the system, the energy absorption and thus the heating rate is exponentially suppressed as a function
of the drive frequency $\Omega$~\cite{rudner_review}. Moreover, in this regime, by ignoring the detail of scatterings, calculating the optical conductivity for non-interacting systems is reasonable. In experiment, to decrease the heating rate of high driving amplitudes, it is convenient to use ultrashort laser pulses (shorter than the time scale for heating processes) leading to short-lived Floquet phases. \deleted{However, by means of tuning the system parameters in a thin film TI, the amplitude of the drive in which a phase transition occurs happens at intensities low enough to avoid heating. Lower light intensities guarantee lower heating rates}~\cite{dabiri1,dabiri2}.

Experimentally, there are several methods to measure optical conductivity. The optical Hall conductivity is measurable by both multi-probe measurement~\cite{foa2014} and the Faraday rotation where the rotation of a linearly polarized radiation is proportional to $\sigma_{xy}(\omega)$. Furthermore, this quantity can be determined in optical lattices by applying a small potential gradient and evaluating the transverse drift of particles in time of flight measurements~\cite{opt}. The absorption and reflection of a probe light is directly related to the longitudinal optical conductivity $\sigma_{xx}$. \added{The occupation of Floquet states is an important subject of the field. Different models have been proposed for the population, however as we show in this work, each one strongly affects optical conductivity. For example, the {\it ac} optical conductivity in driven systems demonstrates anomalous behaviors such as negative longitudinal optical conductivity which is absent in generic static systems. Regarding to population mechanism in these materials, the behavior of all related physical properties such as optical conductivity still needs to be studied in more details. } 

In this paper, we \added{represent a suitable and unified Floquet-Kubo formalism for calculating optical conductivity by expanding the formula appeared already in Ref.~\cite{oka2009}. By means of this formula which is demonstrated to be compatible with others in literature, we} calculate the time-averaged optical conductivity of a lattice model of two dimensional TI such as the model that has been considered in semiconductor quantum wells~\cite{BHZ} \added{or its simulation in cold atoms \cite{coldBHZ}}, irradiated by circular polarized light. Two regimes of the drive are considered: off-resonant and on-resonant regimes. In the high frequency regime, the effective Floquet Hamiltonian is derived and its topological phase diagram is analytically
investigated. \added{To control heating and enhance persistence of the Floquet states, it is worthy to find those topological phases which appear at low intensities and high frequency regimes.} At a medium frequency regime, the topological feature of the system is characterized by direct diagonalizing of the Floquet Hamiltonian in Fourier space giving rise to anomalous edge states. The existence of these edge states are confirmed by calculating the Floquet band structure of the nanoribbon version of Hamiltonian and also the DC optical Hall conductivity. 

To calculate optical response, two models are considered for the occupation of the Floquet states: the ideal and also the mean-energy occupation. By taking the former, quantization of DC optical Hall conductivity is preserved while it is demonstrated that such quantization is absent by taking the latter occupation model. The role of van-Hove singularities appearing in the time-averaged density of states is tracked back into all components of optical conductivity. The optical weight which is not present in the static systems also plays a crucial role to determine the strength of each optical transition. \added{Moreover, a negative optical conductivity which is not observed in {\it ac} optical conductivity of generic static systems is observed in $Re[\sigma_{xx}]$. In fact, the inverted population arising from the turning-on protocole of the driving field is the origin of this behavior. As a result, in this case, there is a stimulated emission of those electrons with population inversion of the Floquet states giving rise to intensification of the probe field. } 

The rest of this paper is organized as follows: In Sec.~\ref{S2} we introduce the tight-binding static model and its special limit at the $\Gamma$ point which results in a typical Hamiltonian of TI thin films in materials such as $Bi_2Se_3$. A brief review on the Floquet theory is presented in Sec.~\ref{S3}. In Sec.~\ref{S4} we derive the Floquet Hamiltonian in two regimes and finally we present the results of the optical conductivity in two frequency regimes and two
occupation models in Sec.~\ref{S5}.

\section{Model}\label{S2}
\begin{figure}
\includegraphics[width= 0.56\linewidth]{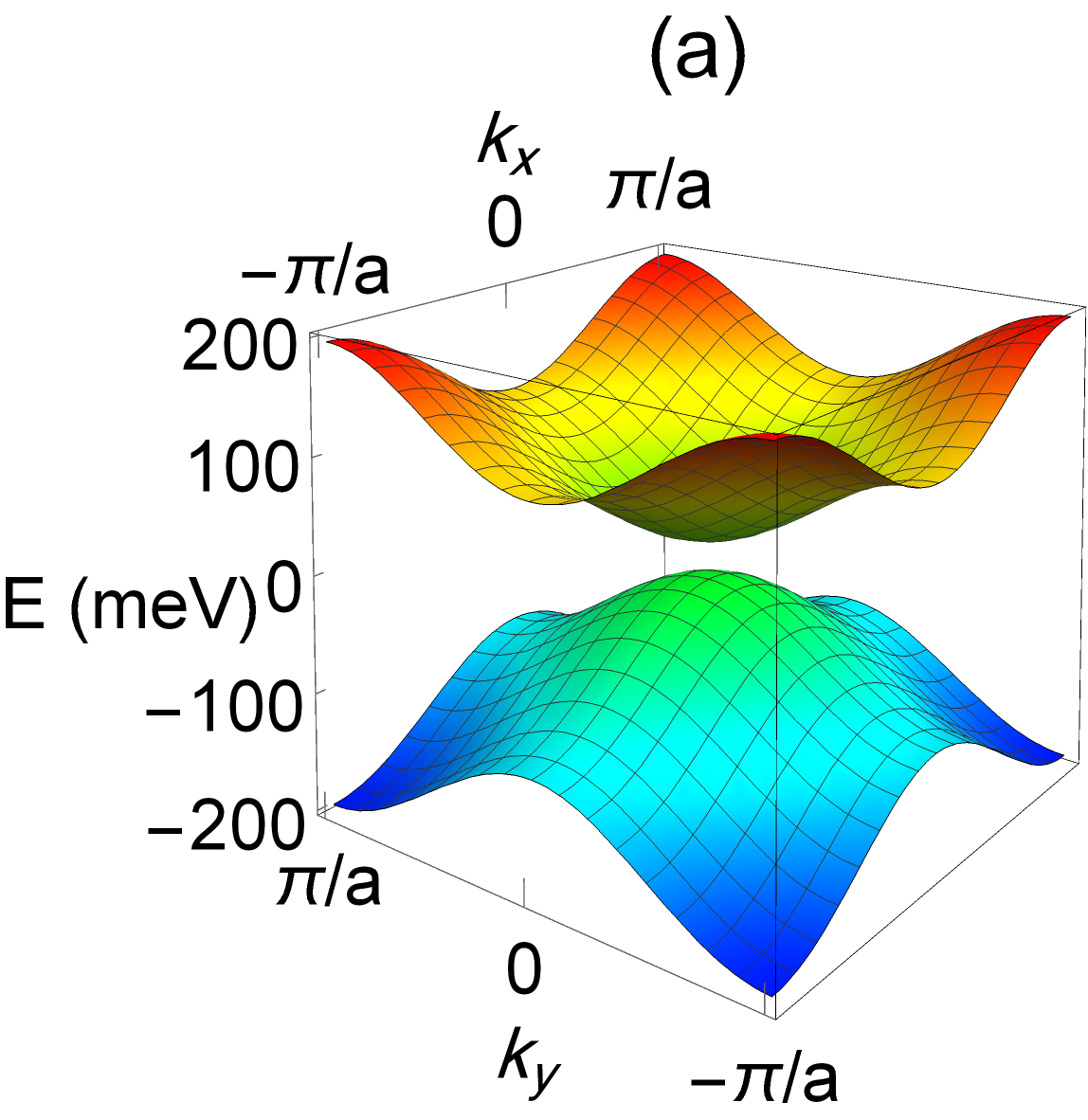}\includegraphics[width=0.44 \linewidth]{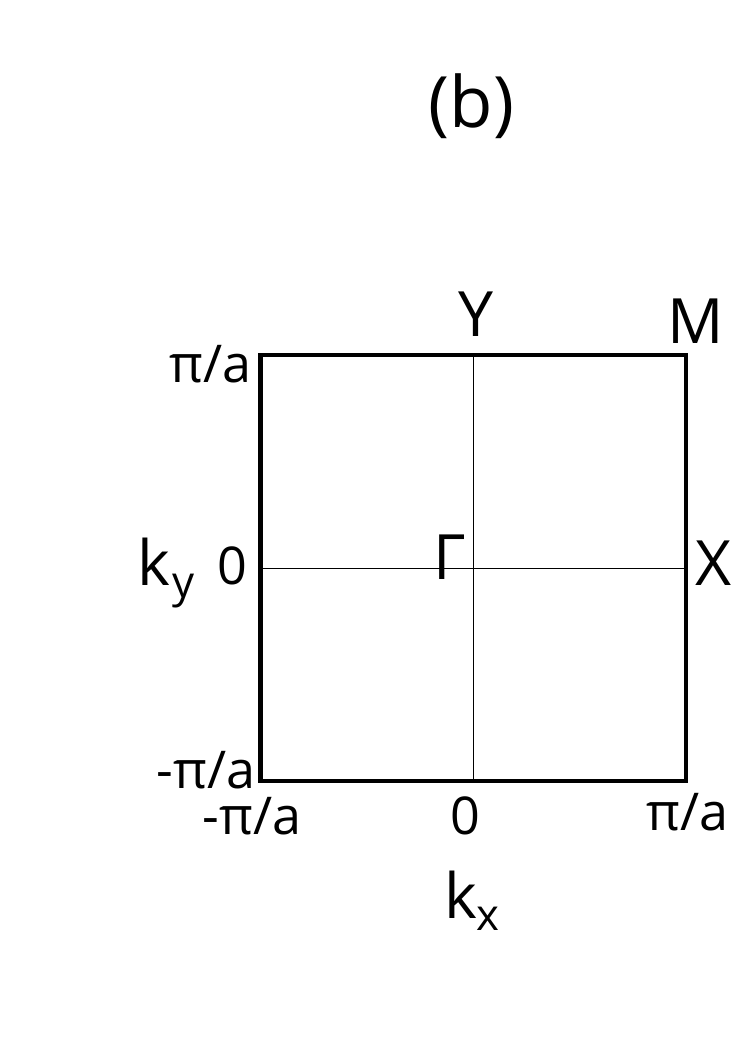} 
\caption{(a) The dispersion relation of 2D TI Hamiltonian defined in Eq.~\ref{hamilk} for $\alpha_z=+1$ and tunneling parameters as $\Delta_0=20$~meV, $\Delta_1=55$~eV\AA$^2$. The panel (b) shows the Brillouin zone of the square lattice of 2D TI and its \deleted{high symmetry} \added{gap closing} points.}
\label{bz}
\end{figure}

%The \emph{tight binding Hamiltonian:}
We begin with a tight binding Hamiltonian on a square lattice introduced for 2D TI such as a quantum well in semiconductor heterojunctions~\cite{BHZ} \added{or its realization in cold atomic systems~\cite{coldBHZ}}. 
Setting $e=\hbar=1$, the Hamiltonian in the momentum space reads
\begin{equation}
\begin{aligned} 
H_\az^{\text{dark}}(\textbf{k})&=\frac{ v_f}{a} \left[\sin{(k_y a)} \sigma _x- \az \sin{(k_x a)}\sigma _y\right]+\Delta_0\sigma_z\\
&+\frac{\Delta_1}{a^2}[4-2 \cos{(k_x a)} -2 \cos{(k_y a)} ]\sigma _z
\label{hamilk} 
\end{aligned} 
\end{equation}
where $a$ is the lattice parameter. 
One gets the low energy effective Hamiltonian for TI thin films,
which is applicable to Bi$_2$Se$_3$ and the (Bi,Sb)$_2$Te$_3$ family materials~\cite{effective,universality,geometrical,PRL2013_zhang,dabiri1,dabiri2},
simply as the $k a\rightarrow0$ limit of Eq.~\ref{hamilk}, namely 
\begin{equation}
\centering
h_\az^{\text{dark}}(\textbf{k})= v_f \left(k_y \sigma _x- \az k_x \sigma _y\right)+\Delta(\textbf{k})\sigma _z
\label{eq:hamilchiral} 
\end{equation}
where $\az=\pm$ is the pseudospin index and $\sigma_i$ are Pauli matrices in the spin-orbital basis. 
Throughout this paper, we assume the Fermi velocity to be $\hbar v_f=297$~meVnm and the lattice parameter as $a=5$~nm.
The first term corresponds to two Dirac cones while $\Delta(\textbf{k})=\Delta_0+\Delta_1 \textbf{k}^2$ comes from the tunneling between two surfaces (when the film thickness is lesser than 5 nm) and opens a gap through a mass term appeared in Dirac cones. 
Two sets of hopping parameters are considered in the following: {\it case (i) } ($\Delta_0=20$~meV, $\Delta_1=55$~eV\AA$^2$), {\it case (ii)} ($\Delta_0=20$~ meV, $\Delta_1=-55$~eV\AA$^2$).
The band structure obtained by directly
diagonalizing the $k$-space Hamiltonian presented in Eq.~\ref{hamilk}, % leads to the dispersion of the system which 
is shown in Fig.~\ref{bz}(a) for $\alpha_z=+1$ and the case (i). 
The first Brillouin zone of the model on a square lattice is demonstrated in Fig.~\ref{bz}~(b) with some \deleted{high symmetry} points indicated on it.

Equation~\ref{hamilk} can be rewritten as 
\begin{equation}
\label{H_k}
H^{dark}(\textbf{k})=T_0+T_x e^{ik_xa}+T_x^\dagger e^{-ik_x a}+T_y e^{ik_ya}+T_y^\dagger e^{-ik_y a},
\end{equation} 
where %the hopping matrices including 
the onsite matrix $T_0$ and the hopping matrices along the $x,y$ directions $T_x$ , $T_y$ are defined as
\begin{equation}
\begin{aligned} 
T_0=&\left(\Delta_0+\frac{4\Delta_1}{a^2}\right)\sigma_z,\\
T_x=&-\frac{\Delta_1}{a^2}\sigma_z+i\frac{ v_f}{2a}({\alpha}_z\sigma_y)\\
T_y=&-\frac{\Delta_1}{a^2}\sigma_z-i\frac{ v_f}{2a}(\sigma_x).
\end{aligned}
\end{equation}
Then, 
%To generalize Hamiltonian, 
the real space tight-binding Hamiltonian is deduced from Eq.~\ref{hamilk} as 
\begin{equation}
\label{TB_Hamiltonian}
H^{dark}_\az=\sum_{\bf r}^{}[ {\bf c}_{\bf r}^\dagger T_0{\bf c}_{\bf r}+(
{\bf c}^\dagger_{\bf r+\hat{x}} T_x{\bf c}_{\bf r}+{\bf c}^\dagger_{\bf r+\hat{y}} T_y{\bf c}_{\bf r}+h.c.)]
\end{equation} 
where ${\bf c}_{\bf r}^\dagger$ and ${\bf c}_{\bf r}$ are the creation and annihilation operators of electron at site $\bf{r}$. 
%Diagonalizing the k-space Hamiltonian presented in Eq.~\ref{hamilk} leads to the dispersion of the system which is shown in Fig. \ref{bz}(a) for $\alpha_z=+1$ and the case (i). The first Brillouin zone of the model has a square shape and is demonstrated in Fig.~\ref{bz}~(b) highlighting some high symmetry points.
%%%The band structure obtained by directly
%%%diagonalizing the $k$-space Hamiltonian presented in Eq.~\ref{hamilk}, % leads to the dispersion of the system which 
%%%is shown in Fig.~\ref{bz}(a) for $\alpha_z=+1$ and the case (i). 
%%%The first Brillouin zone of the model square lattice is demonstrated in Fig.~\ref{bz}~(b) with some high symmetry points indicated on it.

To find the Chern number of Hamiltonian, Eq.~\ref{hamilk}, let us re-write Hamiltonian as $H_\az^{\text{dark}}=\textbf{h}.\boldsymbol{\sigma}$ where $\boldsymbol{\sigma}=(\sigma_x,\sigma_y,\sigma_z)$
whose eigenvalues %of this Hamiltonian is given by $(-\sqrt{h_x^2+h_y^2+h_z^2},\sqrt{h_x^2+h_y^2+h_z^2})$. 
$\pm \sqrt{h_x^2+h_y^2+h_z^2}$.
So the gap is closed only if $h_x=h_y=h_z=0$. If enforcing $h_x=h_y=0$, we can nullify $h_z$ by tuning $\Delta_0,\Delta_1$. Using the above argument, we deduce that the gap closings occur only at \deleted{high symmetry} points ${\Gamma=(0,0), X=(\pi/a,0),Y=(0,\pi/a), M=(\pi/a,\pi/a)}$ albeit each with a different mass terms ${\Delta_0,\Delta_0+4\Delta_1/a^2,\Delta_0+4\Delta_1/a^2,\Delta_0+8\Delta_1/a^2}$, respectively. As explained in Ref.~\onlinecite{geometrical}, the total Chern number is determined by 
$$\added{Ch}=\frac{-1}{2} \sum_{{\bf k} \in {\textbf{D}_i}} \text{sgn}[(\partial_{k_x}\textbf{h}\times \partial_{k_y}\textbf{h})_z] \text{sgn}[h_z],$$ where $\text{sgn}$ is the sign function
and the summation runs over $\textbf{D}_i$ the four \added{gap-closing} \deleted{high symmetry} points of the Brillouin zone. 

\deleted{The necessary condition to have non-trivial topological phases is that the mass terms on each Dirac point have different
signs to guarantee warping of $\textbf{h}$ through the surface Brillouin zone. 
For $\Delta_0 \Delta_1 >0$, all mass terms have the same sign and therefore the system is trivially gapped with 
{\it zero Chern number} for both pseudo-spin indices. 
However, for the case $\Delta_0 \Delta_1<0$, topological phases emerge in more complicated situations. 
Let us suppose that $\Delta_0$ is {\it always positive}, i.e. $\Delta_1<0$. 
%For $\Delta_0 \in [0,4|\Delta_1|/a^2]$, the Chern number for each pseudo-spin is derived as $-\alpha_z$, 
%while for $\Delta_0 \in [4|\Delta_1|/a^2,8|\Delta_1|/a^2]$, the Chern number is $+\alpha_z$. 
For $0 \leq \Delta_0 \leq 4|\Delta_1|/a^2$, the Chern number for each pseudo-spin is $-\alpha_z$
while it is $+\alpha_z$ for $4|\Delta_1|/a^2 < \Delta_0 \leq 8|\Delta_1|/a^2$. 
These two cases correspond to quantum spin Hall insulators (QSHI). 
The sign of all mass terms are again reverted to be positive for the hopping values $\Delta_0 >8|\Delta_1|/a^2$, which leads to a normal insulator (NI) phase. All these conditions, are summarized into the unified form of the Chern number as~\cite{geometrical}}

Applying this formula for our model yields the following result for the Chern number \cite{geometrical}
\begin{equation}
\begin{aligned}
\label{Chern_number}
\added{Ch}_{\alpha_z}=-\frac{\alpha_z}{2} [ &\text{sgn}(\Delta_0)-2\text{sgn}(\Delta_0+\frac{4\Delta_1}{a^2}) \\
&+\text{sgn}(\Delta_0+\frac{8\Delta_1}{a^2})]
\end{aligned} 
\end{equation} 
%Depending on the pseudo-spin polarization, the {\it chirality} of each Dirac point is defined as $\alpha_z \times \text{sgn}(\cos(k_x a) \cos(k_y a))$ giving rise to the chirality for the Dirac points $(\Gamma,X,Y,M)$ as $(+,-,-,+)\times \alpha_z$, respectively.
Regarding to this definition of the Chern number, the value of hopping parameters considered in the case (ii) are in the range
%$\Delta_0=+20~\text{meV} \in[0,4|\Delta_1|/a^2] \equiv [0,88 meV]$ where the Chern number for each pseudo-spin is as $C_{\pm}=\mp 1$.
$\Delta_0=+20~\text{meV} < 4|\Delta_1|/a^2 = 88$~meV where the Chern number for each pseudo-spin is $Ch_{\pm}=\mp 1$.

%%%%%%%%%%%%%%%%%%%%%%%%%%%%%%%%%%%%%%%% FIG 3 %%%%%%%%%%%%%%%%%%%%%%

%%%%%%%%%%%%%%%%%%%%%%%%%%%%%%%%%%%%%%%%%%%%%%%%%%%%%%%%%%%%%%%%%%%%%%%%%%

\section{Floquet Theory}\label{S3}
Applying a time periodic Hamiltonian $H(t)=H(t+T)$ can be implemented by means of Floquet theory \cite{shirley} where energy is not conserved. \deleted{ So it does not make sense talking about the Hamiltonian, instead we can use the unitary operator which describes the full evolution of the system;}. So it is more convenient to work with the {\it stroboscopic}
evolution of the system, $U(T)=\mathcal{T} \int_0^T\exp[-i H(t) dt]$ (where $\mathcal{T}$ denotes time ordering) rather than time-periodic Hamiltonian. 

\deleted{For the sake of simplicity, we look only at the {\it stroboscopic}
evolution of the system, which is actually at the times that are integer multipliers of the driving period T. Because of the periodicity of the Hamiltonian, it is simply proved that
the time evolution operator for stroboscopic times, which is actually at the times that are integer multipliers of the driving period T is written as $U(nT)=[U(T)]^n$.} \deleted{So it is enough only working with the Floquet unitary operator $U(T)$ to describe the system.} \deleted{Having a unitary operator,} \deleted{It is always possible to find a Hermitian operator named {\it Floquet Hamiltonian} $H_F$, where $U(T)=\exp[-i H_F T]$. It should be noted that eigenvalues of this
Hamiltonian, called quasi-energies, are not directly defined for a realistic system
but what is physically meaningful is eigenvalues of}

The eigenvalues of $U(T)$ are of the form $\exp[-i \varepsilon_{\alpha} T ]$ where $\varepsilon_{\alpha}$ represents the quasienergies and $\alpha$ refers to the band index. As a result, $\varepsilon_{\alpha}$ and $\varepsilon_{\alpha}+n \Omega$ show the same eigenstates of the unitary operator with a driving frequency $\Omega=2\pi/T$. 
In other words, in resemblance to the Bloch wave function arising from a periodicity in space, Floquet theorem with a periodicity in time suggests the solution of Schr\"odinger equation to be %as the following form, 
\begin{equation}
\centering
|\psi_{\alpha}(t)\rangle=e^{-i\varepsilon_\alpha t}|\phi_{\alpha}(t)\rangle 
\end{equation}
where the Floquet states $|\phi_{\alpha}(t)\rangle=|\phi_{\alpha}(t+T)\rangle$ are periodic in time. Then the Schr\"odinger equation is rearranged as
\begin{equation}
\centering
[H(t)-i\frac{\partial}{\partial t}] |\phi_{\alpha}(t)\rangle=\varepsilon_\alpha |\phi_{\alpha}(t)\rangle
\end{equation}
where both $H(t)$ and |$\phi_{\alpha}(t) \rangle $, are periodic in time leading to the Fourier version of the above equation. Substituting $H(t)=\sum_m e^{-im\Omega t}H^{(m)}$ and $ |\phi_{\alpha}(t)\rangle=\sum_m e^{-im\Omega t}|\phi^{m}_\alpha\rangle$ with their Fourier expansions, {\it Floquet Hamiltonian} in a matrix form is derived from the following eigenvalue equation,
\begin{equation}
\centering
\varepsilon_\alpha |\phi^{m}_\alpha\rangle=\sum_{m'}[H^{(m-m')}-m\Omega\delta_{mm'}] |\phi^{m'}_\alpha\rangle.
\label{inf}
\end{equation}
Diagonalizing the matrix represented inside the bracket gives us the quasi-energies, $\varepsilon_\alpha$. However, based on periodic property of the quasi-energy, the reduced quasi-energy is defined as $\epsilon_\alpha$ such that it lies in the first Floquet zone $- \Omega/2<\epsilon_\alpha< \Omega/2$. Practically, one should diagonalize Floquet Hamiltonian represented in Eq.~\ref{inf} and find the reduced quasi-energies $\epsilon_\alpha$ and their corresponding eigenvectors and read from them the $|\phi_\alpha^{n}\rangle$. The mean energy of a Floquet band is represented as \cite{wu2011}
\begin{equation}
\centering
\bar{\epsilon}_\alpha=\epsilon_\alpha+\sum_n n\Omega W^n_{\alpha}
\label{menergy}
\end{equation}
where $W^n_{\alpha}=\langle\phi_\alpha^{n}|\phi^{n}_\alpha\rangle$ is the optical weight of each state in the $n$th Floquet replica. Another important quantity which is related to the optical conductivity is the time-averaged density of states (DOS) which is defined as~\cite{wu2011}
\begin{equation}
\centering
\text{DOS}(\omega)=\sum_{n,\alpha} \delta[\omega-(\epsilon_{\alpha}+n\Omega)]W^n_{\alpha}.
\label{dos}
\end{equation}

\section{Driven Hamiltonian}\label{S4}
The model that we are interested in is the 2D TI irradiated by circularly polarized light. In our previous works \cite{dabiri1,dabiri2}, we discussed the effect of circularly \added{and linearly polarized} light on the phase diagram of \added{thin TI, Eq.~ \ref{eq:hamilchiral}, which is a continuous limit of the lattice model represented in Eq.~ \ref{hamilk}. The continuous limit is obtained by just setting $a\rightarrow 0$ in Eq.~ \ref{hamilk}}. Here, we address the effect of \added{circularly polarized light} on optical conductivity of the lattice model.\added{ The time-periodic Hamiltonian is derived by Pierels substitution in $k$-space Hamiltonian (Eq.~\ref{H_k}) by replacing $k_i \longrightarrow k_i+A_i$ which is equivalent to replacing the hopping parameters as $t_{ij} \rightarrow t_{ij} exp(\int_{\mathbf{r}_i}^{\mathbf{r}_j} \mathbf{A}.d\mathbf{l})$ in real space Hamiltonian (Eq.~\ref{TB_Hamiltonian}). Here, the time-periodic vector potential is defined as ${\textbf A}(t)=A_0(\sin\Omega t,\cos\Omega t)$ and $\mathbf{r}_i,\mathbf{r}_j$ represent the positions of the initial and end points, and also $\mathbf{l}=\mathbf{r}_j-\mathbf{r}_i$}. To construct Floquet Hamiltonian represented in the matrix form of Eq.~ \ref{inf}, we need to calculate Fourier components of the time-periodic lattice Hamiltonian $H^{(n)}$,
\begin{equation}
\begin{aligned} 
&H_\az^{(n)}=T_0 \mathcal{J}_n (0) +\mathcal{J}_n(a\mathcal{A})\large[ T_x e^{i k_x a}(-1)^n+T^\dagger_x e^{-i k_x a}\\
&+T_y e^{i k_y a}(i)^{n}+T^\dagger_y e^{-i k_y a}(-i)^{n}\large]\\
&=(\Delta_0+4\frac{\Delta_1}{a^2})\sigma_z \mathcal{J}_n(0)+ \mathcal{J}_n(a\mathcal{A})\large[ \frac{ v_f}{a} ( sin (ak_y+\frac{n\pi}{2}) \sigma_x\\
&-i^{n} sin(a k_x+\frac{n\pi}{2})\az \sigma_y)-\frac{2\Delta_1}{a^2}\sigma_z (i^ncos(a k_x+\frac{n\pi}{2})\\
&+cos (a k_y+\frac{n\pi}{2}))\large]
\label{hf}
\end{aligned} 
\end{equation}
where $\mathcal{A}$ is different with $A_0$ in dimension as $\mathcal{A}\equiv {e A_0/\hbar}$ 
and $\mathcal{J}_n$ is the Bessel Function of the first kind. 
To derive this %the above 
component, we used % the following relation,
\begin{equation}
\frac{1}{T} \int_0^Te^{i a \bf{A}(t).\hat{\theta}}e^{in\Omega t}dt=\mathcal{J}_n(a\mathcal{A}) e^{in(\pi-\theta)}
\label{newhop}
\end{equation}
where the unit vector $\hat{\theta}$ %is the unit vector along the direction which 
connects two nearest neighbor sites.
For our Hamiltonian on a square lattice,
the angle $\theta$ of the unit vector with the $x$-axis
takes only $0,\pi/2,\pi$ and $3\pi/2$.
The resulting Eq.~\ref{hf} reveals that in the presence of circularly polarized light %, to calculate $H^{(n)}$
it is sufficient to multiply the hopping energies between nearest neighbour sites by $\mathcal{J}_n(a\mathcal{A}) e^{in(\pi-\theta)}$.

On the other hand, the current operator is also periodic in time so that $j_u(t)=\partial_{k_u} H(t)=\sum_n e^{-in\Omega t}j_u^{(n)}$ the Fourier components of which are evaluated by Eq.~\ref{hf} as 
$ j_u^{(n)}=\partial_{k_u} H^{(n)}$, namely 
\begin{equation}
\begin{aligned} 
&{\bf J_\az^{(n)}}=\mathcal{J}_n(a\mathcal{A})\large[i a ( T_x e^{i k_x a}(-1)^n- T^\dagger_x e^{-i k_x a}) \hat{x}\\
&+i a ( T_y e^{i k_y a}(i)^{n}- T^\dagger_y e^{-i k_y a}(-i)^{n})\hat{y} \large]\\
&= \mathcal{J}_n(a\mathcal{A})\large[ v_f ( cos (a k_y+\frac{n\pi}{2}) \sigma_x \hat{y}\\
&-i^{n} cos(a k_x+\frac{n\pi}{2})\az \sigma_y\hat{x})+\frac{2\Delta_1}{a}\sigma_z (i^n sin(a k_x+\frac{n\pi}{2})\hat{x}\\
&+sin (a k_y+\frac{n\pi}{2})\hat{y})\large]
\label{current_fourier}
\end{aligned} 
\end{equation}
where the index $u$ runs over $x$ and $y$ directions. 
From now on, we set the pseudospin index $\az=+1$ and drop this index in all calculations. 
The case of $\az=-1$ can be similarly considered.

Now that we have the new hopping parameters corresponding to $H^{(n)}$ in Eqs.~\ref{newhop} and \ref{hf}, 
it is easy to construct the Floquet matrix of Eq.~\ref{inf}. The matrix has infinite dimensions.
However one can truncate it at some dimension which guarantees the convergence of results. 

%%%%%%%%%%%%%%%%%%%%%%%%%%%%%%%%%%%%%%%% FIG 3 %%%%%%%%%%%%%%%%%%%%%%
\begin{figure}
\includegraphics[width= \linewidth]{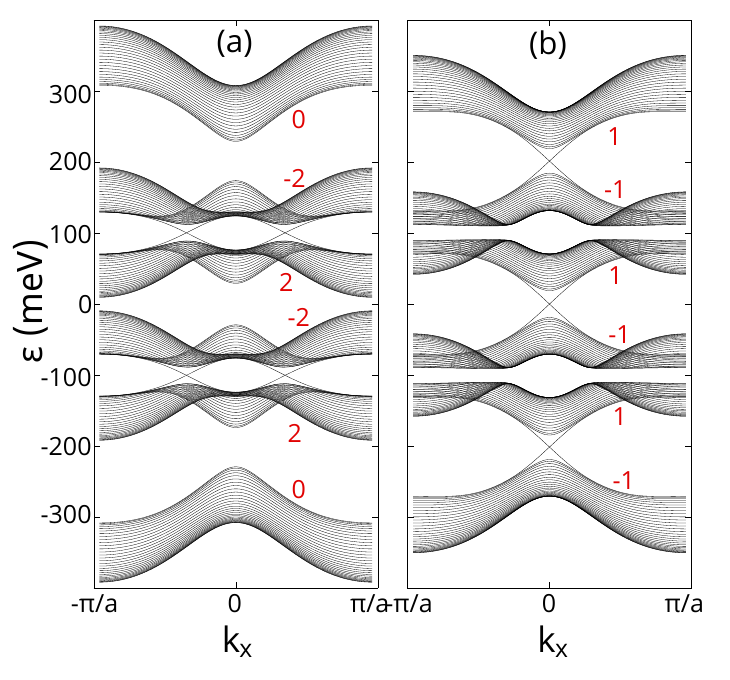} 
\caption{The spectrum of the nanoribbon version of Floquet Hamiltonian represented in Eq.~ \ref{inf}. The panel {\it a} ({\it b}) belongs to the tunneling parameters taken from the {\it case (i)} ({\it case (ii)}) which is primary in the {\it NI} ({\it Chern insulator}) phase. The frequency of drive is $\Omega=200$~meV and its intensity is considered to be as $\mathcal{A}=0.1~\text{nm}^{-1}$. The red integers \added{show the Chern number of the corresponding bulk band}.}
\label{tworib}
\end{figure}

%%%%%%%%%%%%%%%%%%%%%%%%%%%%%%%%%%%%%%%%%%%%%%%%%%%%%%%%%%%%%%%%%%%%%%%%%%
The dispersion of the nanoribbon version of Floquet Hamiltonian is shown in Figs.~\ref{tworib} (a,b). The spectrum is truncated with $n=0,\pm1$ subbands. The considered drive frequency $\Omega=200$~meV is lower than the bandwidth. The bandwidth is approximately 0.3- 0.4~eV (see Fig.~\ref{om1000} (a),(d)) and regarding to the given driving frequency, only one photon-assisted processes are represented in the spectrum. \added{Figs.}~\ref{tworib}(a,b) correspond to the case (i) and (ii), respectively.
The Chern number of \added{the corresponding bulk} bands, represented by the red integers, can be derived using the Fukui-Hatsugai-Susuki method~\cite{Discrete_BZ}. This technique uses the eigenvectors of Floquet Hamiltonian attributed to each band. It is evident in the spectrum shown in Fig.~\ref{tworib} that the Chern number is also equal to the difference between the number of \added{chiral} edge modes coming from the above and exiting from the bottom of each band as noted in Ref.~\onlinecite{rudner}. \added{Note that in this rule, the sign of the Chern number appears as the chirality of an edge mode, e. g. the chirality of the edge modes displayed in Fig.~\ref{tworib}(a) is opposite of the chirality of the edge modes emerging in Fig.~\ref{tworib}(b)} \cite{rudner}.

As seen in Fig.~\ref{tworib} (a), on-resonant driving field applied on the primary
normal insulating phase can induce two edge states inside the dynamical gap called   \emph{anomalous edge states}. On the other hand, for the case (ii), Fig.~\ref{tworib} (b) implies that the normal insulating phase persists in the Chern insulator induced by the right-handed polarized light albeit at low intensities. \added{To have complete view,} the phase diagram of an irradiated 2D insulator is drawn in Fig.~\ref{2phase} (b) as a function of the amplitude $\mathcal{A}$ for the case (i) and at a frequency, $200$~meV lower than the band width. \added{In this figure, the minimum and maximum boundaries of the upper Floquet band in the central Floquet zone are indicated with the dotted lines while the blue region between these dotted lines shows the width of this band. It was checked that once the dynamical gap is closed, the Chern number of each band (the red integers) is changed.}
\subsubsection{Effective Hamiltonian at High Frequency Regime}
In high frequency regime (larger than the band width) and low amplitudes of the drive, the full Floquet Hamiltonian can be projected onto the central Floquet zone %appearing in 
i.e. the interval $(-\Omega/2,\Omega/2)$. This piece of band structure has the most contribution to the time-averaged DOS and other physical properties such as optical conductivity compared to the other sidebands; $\langle\phi^0_\alpha|\phi^0_\alpha\rangle \gg \langle\phi^n_\alpha|\phi^n_\alpha\rangle$ where $n\neq 0$.

\added{By using perturbation theory, it is shown that for off-resonant drive $|\phi^n\rangle=\frac{H^{(n)}}{n\Omega}|\phi^0\rangle~~\text{for}~n\neq0$ \cite{kitagawa}. We assume the dimensionless parameter $\frac{\mathcal{A}v_f}{\Omega}\ll1$ to be in the weak driving regime and just the central side band has the most effect}.

So, it is straightforward to consider the effect of sidebands on the central bands, perturbatively in this regime. This leads us to an effective Floquet Hamiltonian which describes low intensity driving fields. As a result, one can use a series expansion in powers of inverse frequency to find out this Hamiltonian~\cite{Bukov,eckart,bw} as the following
\begin{equation}
H^{\text{eff}}=H^{(0)}+\sum_{n}(n \Omega)^{-1}[H^{(-n)},H^{(+n)}]+O(1/(\Omega)^2)
\label{photon_H}
\end{equation}
where the Fourier components of time-dependent Hamiltonian are defined as $H^{n}=1/T \int_{0}^{T} H(t) e^{ in |\Omega| t} dt$. The effective two band Hamiltonian is easily calculated by means of the Fourier components given in Eq.~\ref{hf} and then the optical conductivity is evaluated by using the static Kubo Formula instead of the Floquet-Kubo version which will be presented in Eq.~\ref{exkubo}. We checked that the results of optical conductivity extracted from the full driven Floquet Hamiltonian and also the effective Floquet Hamiltonian are very compatible with each other as long as the off-resonant regime is respected. If the strength of light becomes stronger, then the higher orders of expansion \added{i.e. $O(\Omega^{-2}),O(\Omega^{-3}),...$} in Eq.~\ref{photon_H} are required. \added{One can check the validity of the truncated expansion numerically by comparing the band structure of the effective Hamiltonian (Eq.~\ref{photon_H}) and the Floquet Hamiltonian Eq.~\ref{inf}.}

In this stage, let us try to find out the phase diagram of system in the off-resonant regime and at low intensity of irradiated light by using the effective Hamiltonian which is derived by substituting Eq.~\ref{hf} into Eq.~\ref{photon_H}. The result can be expressed in the following form, $H_{\text{eff}}=\textbf{h}^{\text{eff}}.\boldsymbol{\sigma}$, defining $C=a^2\mathcal{J}_0(a\mathcal{A})$ and $D=8\frac{\Delta_1}{\Omega}\mathcal{J}^2_1(a\mathcal{A})$, we have

\begin{equation}
\begin{aligned}
h_x^{\text{eff}}&=\frac{v_f}{a^3}[C sin(a k_y)-D cos(ak_x)sin(ak_y)]\\
h_y^{\text{eff}}&=\frac{v_f}{a^3}[-C sin(a k_x)+D sin(ak_x)cos(ak_y)]\\
h_z^{\text{eff}}&=\Delta_0+4\frac{\Delta_1}{a^2}+\frac{4v_f^2\mathcal{J}^2_1(a\mathcal{A}) cos (ak_x) cos (a k_y)}{a^2\Omega}\\
&+\frac{-2 \Delta_1 \Omega \mathcal{J}_0(a\mathcal{A}) [cos (ak_x) +cos (a k_y)]}{a^2\Omega}
\label{heff}
\end{aligned}
\end{equation}

\begin{figure}
\includegraphics[width= \linewidth]{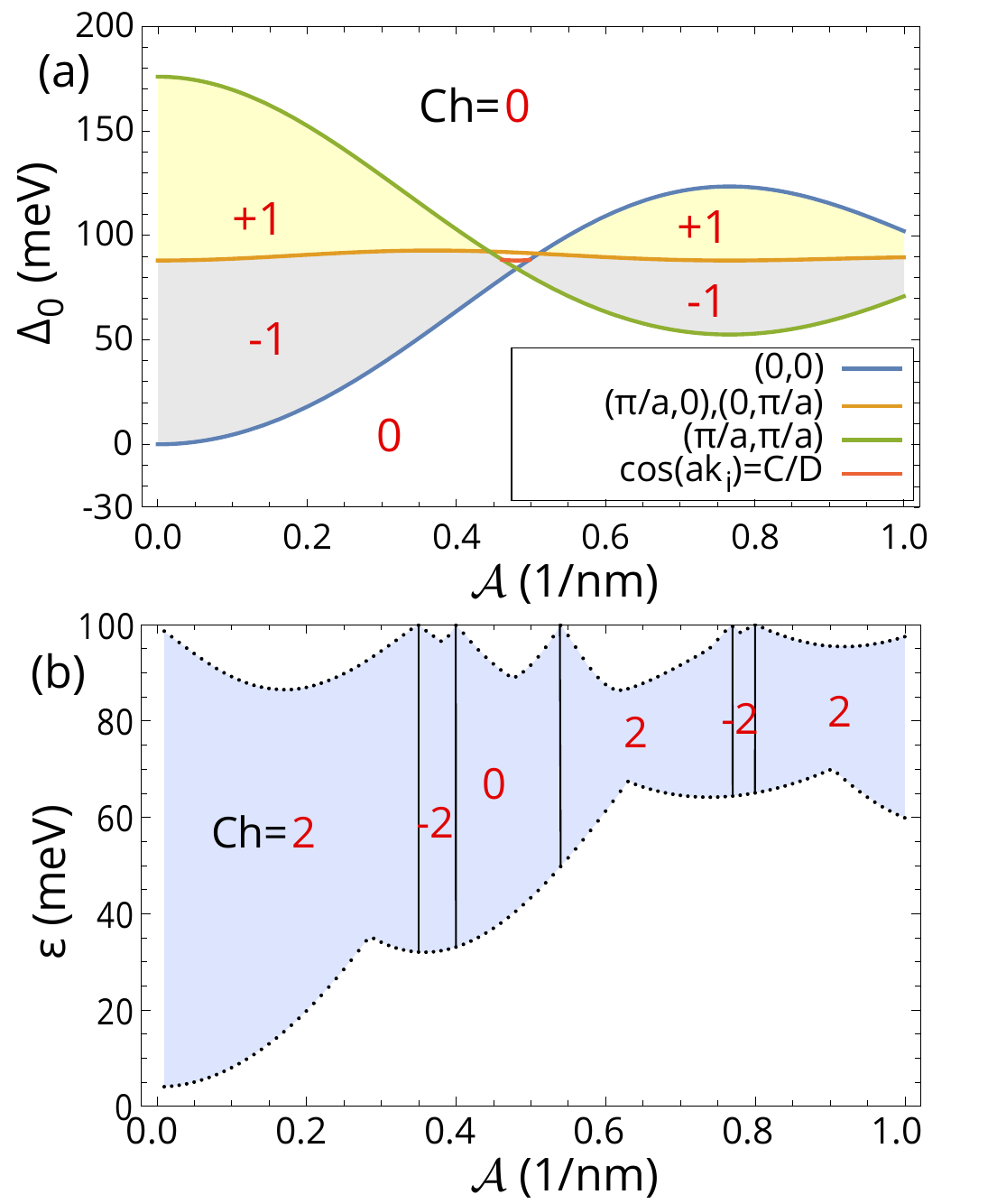} 
\caption{(a) Phase diagram of irradiated model for $\Delta_1=-55$~eV\AA$^2$ and the frequency of $\Omega=1000 $~meV as a function of $\Delta_0$ and $\mathcal{A}$. The chern number is derived by using the effective Hamiltonian defined in Eq.~\ref{heff}. The red integers show the lower band Chern number. The border lines are plotted for the gap closing \added{points} \deleted{conditions at high symmetry points} and the other conditions explained in the text. (b) The phase diagram as a function of $\mathcal{A}$ for the tunneling parameters chosen from the case (i) and driven by a light with the frequency of $\Omega=200$meV. The minimum and maximum of the upper Floquet band are shown by the dots. The red integers show the upper Floquet band Chern numbers. Everywhere that the dynamical gap is closed, the Chern number is changed.}
\label{2phase}
\end{figure}

Since the other terms are negligible for small amplitudes of the drive, we have included just $n=0,\pm1$ terms in Eq.~\ref{photon_H}. The gap is closed if %and only if 
$h^{\text{eff}}_x=h^{\text{eff}}_y=h^{\text{eff}}_z=0$. Once $h^{\text{eff}}_x=h^{\text{eff}}_y=0$, the term $h^{\text{eff}}_z$ can also be set to zero by tuning $\Delta_0$ and $\Delta_1$. So the following two conditions must be simultaneously satisfied for the gap closing points,
\begin{equation}
\left \{
\begin{aligned}
&sin(ak_y)[C -D cos(ak_x)]=0\\
&-sin(ak_x)[C-D cos(ak_y)]=0\\
\end{aligned}\right.
\end{equation}
Indeed the points $\Gamma,X,Y,M$ are the first solution of the above equations where $sin(ak_x)=sin(ak_y)=0$. As long as $|C|=|D|$, there are two conditions as $sin(ak_y)=0$ or $sin(ak_x)=0$ which counts the other solutions. The third solution occurs at $cos(ak_x)=cos(ak_y)=C/D$. These are a complete set of solutions, in each case we set $h^{\text{eff}}_z=0$ to find the border lines of phases. The resultant phase diagram is shown in Fig.~\ref{2phase} (a) where we have set $\Delta_1=-55$~eV\AA$^2$.
It is obvious that those $\Delta_0$ intervals in which the Chern number of the system is nonzero varies with the light intensity at high frequency regime. 
%%%%%%%%%%%%%%%%%%%%%%%%%%%%%%%%%%%%%%%% FIG 3 %%%%%%%%%%%%%%%%%%%%%%
%%%%%%%%%%%%%%%%%%%%%%%%%%%%%%%%%%%%%%%%%%%%%%%%%%%%%%%%%%%%%%%%%%%%%%%%%%

\section{Optical Conductivity}\label{S5}

The linear response theory and Kubo formula for optical conductivity of time periodic systems has been already developed in several studies~\cite{oka2009,oka2010,dehghaniOC}. The time-averaged optical conductivity at zero temperature can be derived from 
a compact equation which aggregates both longitudinal and Hall optical conductivity %in a unified formula. 
(its equivalence with those reported in literature is proven in Appendix~\ref{Eqv_Kubo_Floquet})
\begin{equation}
\begin{aligned}
& \sigma^{uv}(\omega)=-i \sum_{\alpha,\gamma>\alpha} \sum_{m} \frac{f_\alpha-f_\gamma}{\epsilon_\alpha-\epsilon_\gamma-m\Omega}\times \\
& \big[ \frac{j_{\alpha\gamma}^{v (m)}j_{\gamma\alpha}^{u (-m)} }{\omega+i0^+-(\epsilon_\alpha 
-\epsilon_\gamma-m\Omega)} +\frac{ j_{\alpha\gamma}^{u (m)} j_{\gamma\alpha}^{v (-m)}}{\omega+i0^++\epsilon_\alpha-\epsilon_\gamma-m\Omega}\big]
\label{exkubo}
\end{aligned}
\end{equation}
where $\omega$ is the probe frequency and $\alpha,\gamma=1,2$ refer to the band indices indicating inter-band transitions.
$f_{\alpha}$ denotes the band occupation $\alpha$. Here $j_{\alpha\gamma}^{u(n)}$ is the matrix elements of the current operator along the $u$-direction. The sum over ``$m$'' refers to the $m$-photon processes. In Appendix~\ref{current_op}, we define the matrix elements of the current operator and show how to expand this formula in terms of Fourier components of Floquet modes for numerical usage.

The above equation can be interpreted based on some excitations from the state $|\phi_\alpha^{0} \rangle$ in the central Floquet zone by means of absorption $m$ photons to the virtual states in the $m$th side band, $|\phi_\gamma^{m} \rangle$ and then coming back to the initial state by emission of those $m$ photons.

\subsection{Off-resonant Regime}
\begin{figure}
\includegraphics[width= \linewidth]{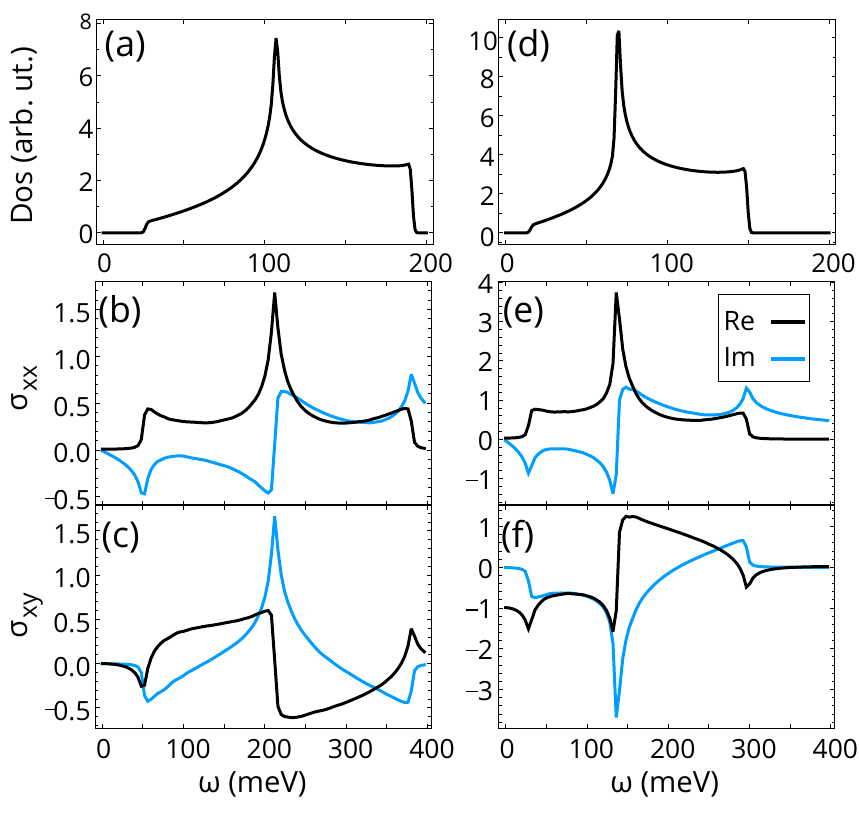} 
\caption{Time-averaged density of states and optical conductivity (in units of $e^2/h$) of the 2D TI irradiated by circularly polarized light in high frequency regime. The frequency of the drive is $\Omega=1000$~meV and the intensity $\mathcal{A}=0.10~\text{nm}^{-1}$. The real and imaginary parts of optical conductivity are displayed by different colors. The panels {\it a-c} ({\it d-f}) are belonging to the tunneling parameters chosen from the {\it case (i)} ({\it case (ii)}) which are primary in the {\it NI} ({\it Chern insulator}) phase.}
\label{om1000}
\end{figure}

The density of states and the real and imaginary parts of time-averaged optical conductivity of irradiated 2D TI with frequency $\Omega=1000$~meV calculated by Eqs.~\ref{dos} and \ref{exkubo}
are shown in Fig.~\ref{om1000}.
The frequency of light is larger than the bandwidth of the system, so \emph{off-resonant} regime is satisfied. Here we have two bands where the lower(upper) Floquet band is completely filled (empty). So in Eq.\ref{exkubo}, the occupations are $f_{(\alpha=1)}=1 , f_{(\alpha=2)}=0$. We call this type of occupation the ``{\it ideal occupation}''.

Figures~\ref{om1000} (a-c) and (d-f) are plotted for the case (i) and (ii) and correspond to two different values of hopping parameters $\Delta_0$ and $\Delta_1$ which are primary in NI and Chern insulator phase, respectively. In Fig.~\ref{om1000}(b),(e) the real and imaginary part of longitudinal optical conductivity are presented in which their dips and peaks correspondingly follow van-Hove singularities appeared in the time-averaged density of states displayed in Fig.~\ref{om1000} (a), (d). Regarding to inter-band transitions, the position of dips and peaks in optical conductivities must emerge exactly at twice of that frequency that van-Hove singularity appears in the density of states. \added{To explain more, the optical weight of Floquet eigenstates are decaying rapidly for the sidebands of $(|n| \gg \mathcal{W}/\Omega)$, where $\mathcal{W}$ is the band width. For sufficiently large driving frequency and weak intensity (assuming dimensionless parameter $\frac{\mathcal{A}v_f}{\Omega}\ll1$ ), the Floquet states are localized in the central Floquet bands, $n=0$. Therefore, the significant transitions mostly occur among these two central bands. Regarding the electron-hole symmetry of Hamiltonian, the photon-dressed bands are also symmetric around zero energy. Then we expect that if the density of states has a peak (or sudden change) at a given energy, let's say $E_{p}$, optical conductivity shows an extremum structure at an energy $2 E_{p}$ because more transitions (or a sudden change in the number of transitions) occur between the lower band at energy $\epsilon\approx-E_{p}$ and the upper band at energy $\epsilon\approx E_{p}$} 

$Re[\sigma_{xx}]$ is positive for all ranges of probe frequencies which originate from the occupation of states in resemblance to the static ones. Any peak and dip in the real and imaginary parts of optical Hall conductivity, Fig.~\ref{om1000} (c), (f) also corresponds to van-Hove singularities in density of states. At zero probe frequency, $\omega=0$, $Re[\sigma_{xy}]$ is quantized and \deleted{goes}\added{proportional} to the value of the Chern number of the lower Floquet band. Regarding \added{to} the phase diagram represented in Fig.~\ref{2phase} (a), for the amplitude of the drive setting to $\mathcal{A}=0.1~ \text{nm}^{-1}$ and the hopping energy between two surfaces as, $\Delta_0=20$~meV, the Chern number is equal to -1 for the case (ii) and zero for the case (i).

\deleted{As stated in Sec.~\ref{S4}, in the off-resonant regime, one can calculate driven optical conductivity by means of static Kubo formula provided that instead of full driven Hamiltonian, we use the effective static Hamiltonian which has a series expansion as shown in Eq.~\ref{photon_H}. The result of optical conductivity are the same as what is shown in Fig.~\ref{om1000}.}

%%%%%%%%%%%%%%%%%%%%%%%%%%%%%%%%%%%%%%%% FIG 3 %%%%%%%%%%%%%%%%%%%%%%

%%%%%%%%%%%%%%%%%%%%%%%%%%%%%%%%%%%%%%%%%%%%%%%%%%%%%%%%%%%%%%%%%%%%%%%%%%

%%%%%%%%%%%%%%%%%%%%%%%%%%%%%%%%%%%%%%%% FIG 3 %%%%%%%%%%%%%%%%%%%%%%
\begin{figure}
\includegraphics[width= \linewidth]{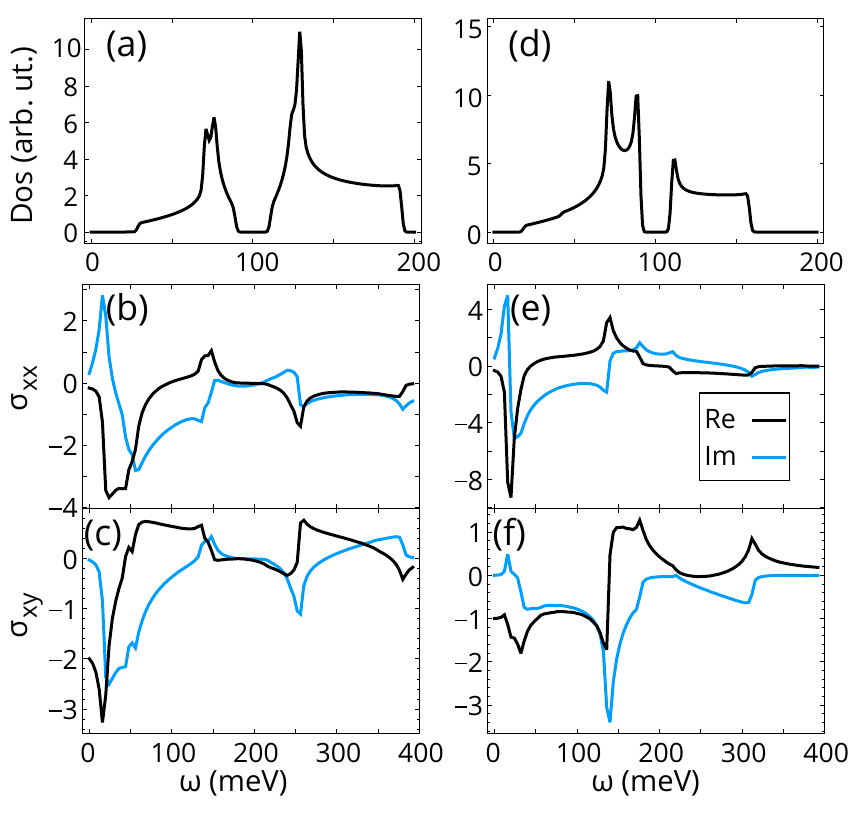} 
\caption{Time-averaged density of states and optical conductivity (in units of $e^2/h$) of 2D topological insulator irradiated by on-resonant circularly polarized light. The frequency of the drive is $\Omega=200$~meV and the intensity is $\mathcal{A}= 0.1~\text{nm}^{-1}$. Real and imaginary parts of optical conductivity are displayed by different colors. For this calculation, the \emph{Ideal occupation} model is assumed. The panels {\it a-c} ({\it d-f}) belong to the tunneling parameters chosen from the case (i) (the case (ii)) which are primary in the NI (Chern insulator) phase.}
\label{om200}
\end{figure}

%%%%%%%%%%%%%%%%%%%%%%%%%%%%%%%%%%%%%%%%%%%%%%%%%%%%%%%%%%%%%%%%%%%%%%%%%%
\subsection{On-resonant Regime} 
Now we study optical conductivity of a 2D TI irradiated by light \added{in medium frequency regime by supposing two types of the occupation for the Floquet states.} % (Figs. \ref{om200} and \ref{200mean}). First let us introduce two occupation models. It should be mentioned that
The non-equilibrium distribution function for the population of the Floquet states depends strongly on the initial condition of light shining on the sample and also on the relaxation mechanisms such as electron-phonon couplings,   radiative recombination, etc.~\cite{seetharam}. So the Fermi-Dirac distribution as the static case does not necessarily occur in experiment. \added{However, it was shown that a nearly ideal occupation of Floquet states can be achieved by engineering the Bose and Fermi baths coupled to the system \cite{seetharam}}.

In this paper, two occupation models are assumed: \emph{ideal} and \emph{mean energy} occupation.
By the former , we mean that the lower (upper) Floquet bands are completely occupied (unoccupied) while by the latter we mean that the Floquet states on each $k$-point are occupied depending on the mean energy $\bar \epsilon_\alpha$ of that point (Eq.~\ref{menergy}) according to the Heaviside step function $f_ {\alpha}=\theta(-\bar{\epsilon}_\alpha)$. %where $\theta$ is the Heaviside step function. 
Throughout this work, the Fermi energy is fixed at the band center, $E_F=0$. %It is possible to calculate the mean energy of each of two bands in the central Floquet zone by using Eq.~\ref{menergy}. 
The bands in other sidebands will be occupied similar to the central bands. Another occupation type used in literature %~\cite{oka2010} is 
originates from projection of the Fermi-Dirac distribution in non-driven situation on the Floquet basis sets and is called ``sudden approximation''~\cite{oka2010,wu2011}, or ``quench''~\cite{seradjeh2020}. Although mean energy occupation \added{is a phenomenological assumption and }is quite different from the projected distribution for a closed system, the whole feature of optical conductivities calculated by these distributions shows the same behaviors \cite{wu2011}.

It was checked that in contrast to the on-resonant case, in the off-resonant regime, the mean energy of the lower
(higher) Floquet band is negative (positive) for all points of the Brillouin zone \added{which resembles ideal occupation}. Therefore, for the off-resonant drive two mentioned models for the occupation coincide with each other. \added{However, the results arising from these occupation models are totally different if we work in the on-resonant regime.}

\subsubsection{Ideal Occupation }
\begin{figure}
\includegraphics[width= \linewidth]{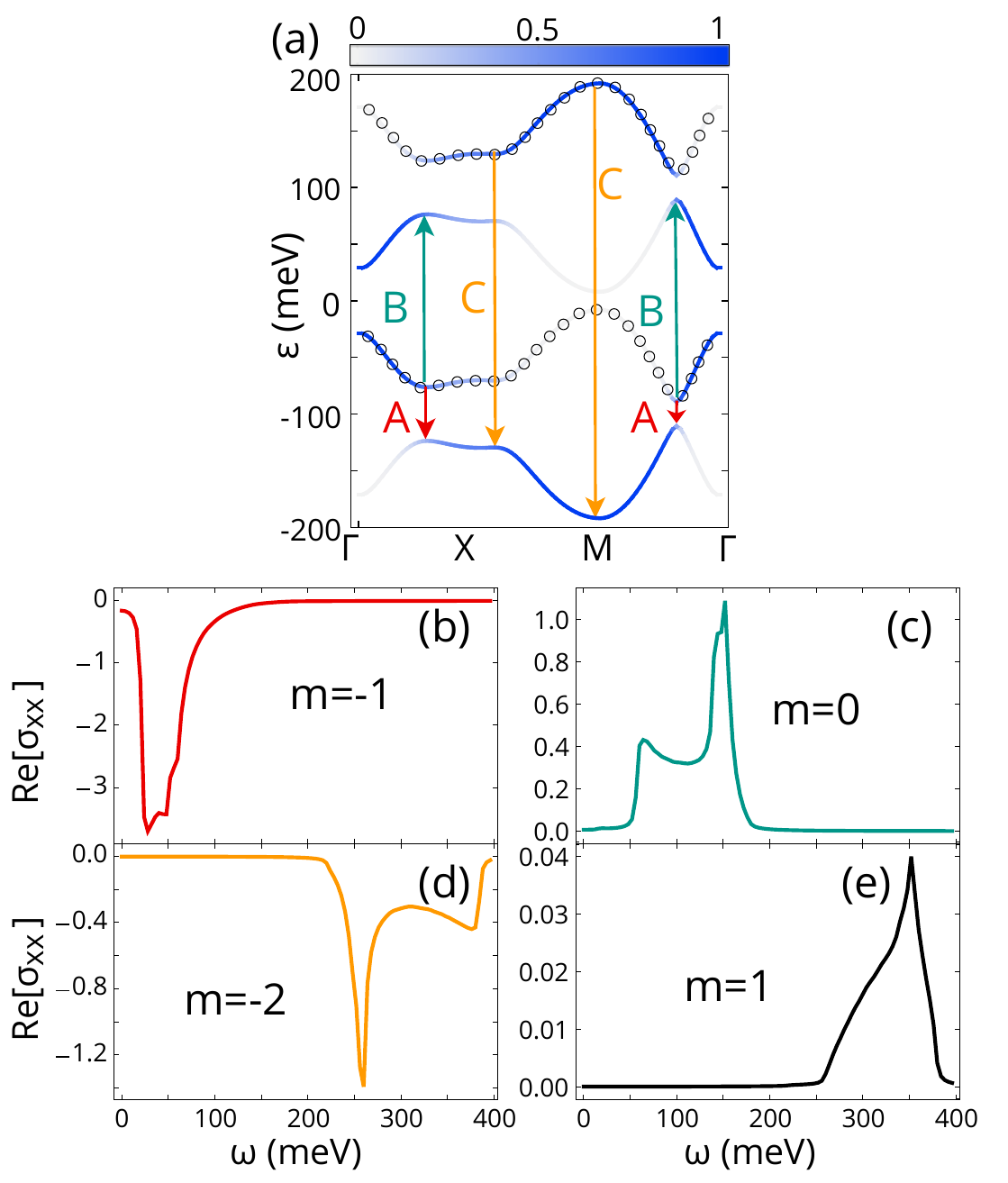} 
\caption{(a) The dispersion relation of Floquet Hamiltonian for the case (i): $\Delta_0=20~\text{meV}$ and $\Delta_1=55$~eV\AA$^2$. The frequency of the drive is $\Omega=200$~meV and its intensity $\mathcal{A}=0.1~\text{nm}^{-1}$. The colored scale bar on each band shows the optical weights $\langle\phi_\alpha^n|\phi_\alpha^n\rangle$. Arrows indicated in panel (a) show some important optical transitions and the circles display the occupied bands. The ideal occupation model is considered in this calculation. (b)-(e) The decomposition of the summation terms for the real part of longitudinal optical conductivity in terms of m-photon processes. Note that the A,B,C transitions in panel (a) correspond to $m=-1,0,-2$ photon processes and the associated conductivities are shown in (b),(c),(d), respectively.}
\label{comp}
\end{figure}

In Fig.~\ref{om200}, we plot the density of state and optical conductivity of 2D TI irradiated by on-resonant light and an assumption of the ideal occupation. The frequency of drive is considered $\Omega=200~$meV which is lower than the band width. The real and imaginary parts of optical conductivity are displayed with different colors \added{in Figs. \ref{om200} (a-c) and (d-f) which are plotted for the case (i) and case (ii), respectively. The dynamical gap appearing in the on-resonant drive occurs at $\Omega/2=100~$meV as shown in the density of states (Figs.~\ref{om200}(a),(d)).} 
In this figure, there are some important physical remarks. First, contrary to the off-resonant regime (Figs.~\ref{om1000}(b,e)) for which the real part of the longitudinal optical conductivity (Re$[\sigma_{xx}]$) is positive at all probe frequencies, in the on-resonant regime with the ideal occupation, this quantity becomes negative in some frequency regions (Figs.~\ref{om200}(b,e)). The second remarkable point is that the emergence of the peaks and dips in optical conductivity occurs at \added{a probe energy twice than the energy at which the van-Hove singularities appear in the density of states (Figs. 
~\ref{om200}(a,d)), of course except for some peaks or dips at low frequencies}. The third remark is focused on the limit of zero probe frequency Re$[\sigma_{xy}(\omega \rightarrow 0)]$ (see Figs.~\ref{om200}(c,f)) which is compatible with the Chern numbers calculated numerically and confirmed by the Floquet band structure of the nanoribbon version (see Figs.~\ref{tworib}(a,b)). Now, we explain the first and second points with the help of Fig.~\ref{comp}(a). 
%%%%%%%%%%%%%%%%%%%%%%%%%%%%%%%%%%%%%%%% FIG 3 %%%%%%%%%%%%%%%%%%%%%%

%%%%%%%%%%%%%%%%%%%%%%%%%%%%%%%%%%%%%%%%%%%%%%%%%%%%%%%%%%%%%%%%%%%%%%%%%%

\begin{figure}
\includegraphics[width= \linewidth]{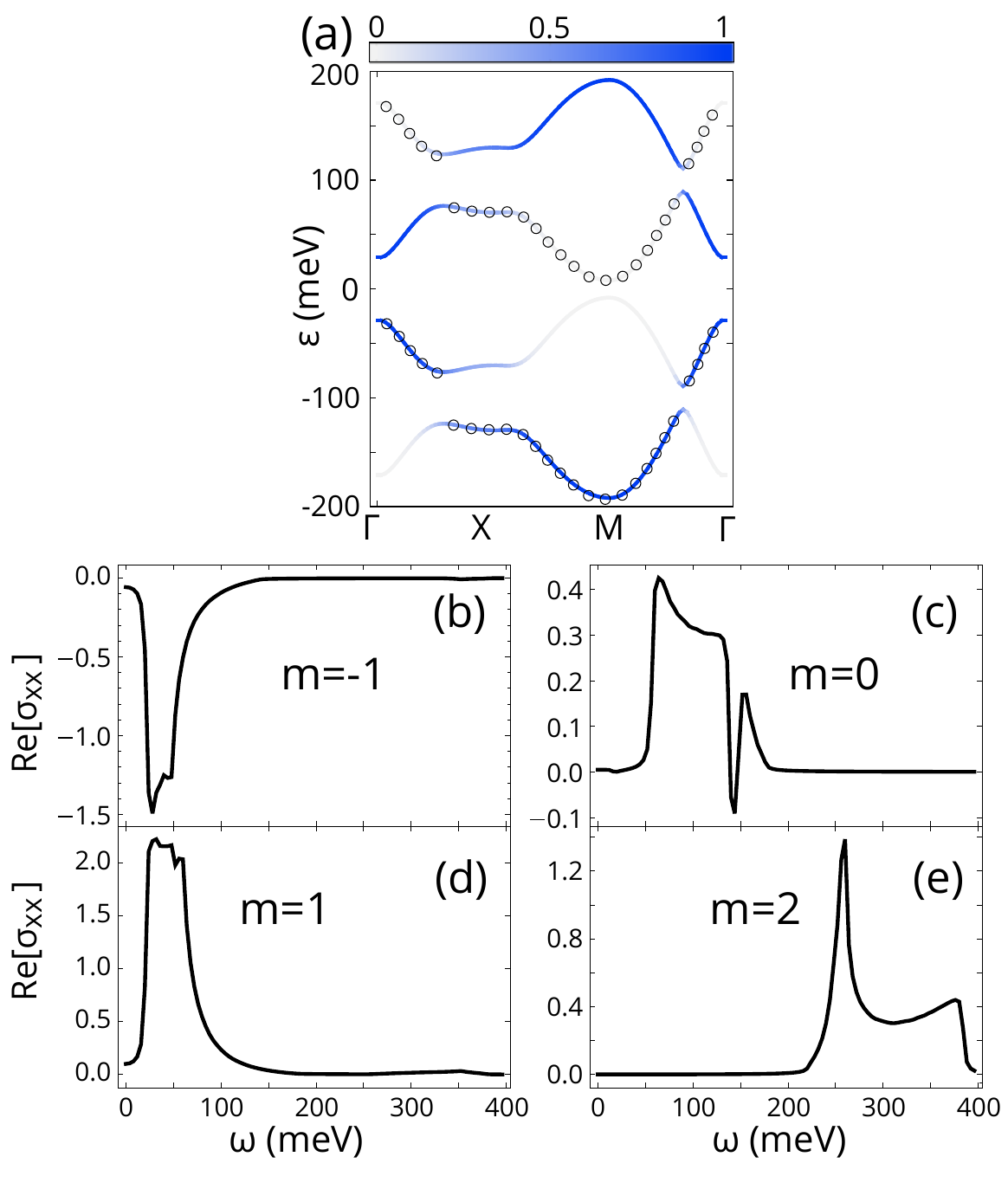} 
\caption{ (a) The dispersion relation of Floquet Hamiltonian for 2D topological insulator. The optical weights are displayed by the colored scale bar and the occupation model is assumed to be the ``\emph{mean-energ}'' occupation. (b)-(d) The decomposition of the real part of the longitudinal optical conductivity in terms of m-photon processes. The tunneling and drive parameters are the same as given in Fig.~\ref{comp}}
\label{phmean}
\end{figure}

{\it Optical transitions}: \added{Suppose that the hopping parameters between two surfaces is given as in the case (i).} Figs.~\ref{comp}(a) and \ref{phmean}(a) represent the Floquet band structure of 2D TI for the ideal and mean-energy occupation, respectively. The hollow circles \added{indicated} on the bands represent the occupation. Assuming the ideal occupation, as shown in Fig.~\ref{comp}(a), the lower band belonging to the central Floquet bands is fully filled while the upper one is empty. The same pattern for the occupation of the states is copied in the Floquet sidebands. On the other
hand, if the mean-energy is used to calculate the occupation for each band, as seen in Fig.~\ref{phmean}(a), the bands would be partially filled. In this section, we assume a drive with moderate frequency $\Omega=200$~meV and also a scaled intensity $\mathcal{A}=0.1~\text{nm}^{-1}$. The color scale on the bands indicates the optical weights of the states defined as $W^n_\alpha=\langle\phi_\alpha^n|\phi_\alpha^n\rangle$. As explicitly seen in Eq.~ \ref{dos}, the weights directly affect the density of states and consequently the optical conductivity. 

The vertical arrows in Fig.~\ref{comp}(a) %Fig.\ref{phmean}(a) 
are as indicators for the most important optical transitions with remarkable traces in the optical conductivity. The A and C transitions have negative contributions to Re$[\sigma_{xx}]$; see Fig.~\ref{om200}(b). \added{From the view point of population inversion of the Floquet band spectrum, negative optical conductivity is reasonable. Population inversion causes an stimulated emission which amplifies the probe field. In other words, since the real part of optical conductivity is directly related to the absorption, a negative sign of this quantity leads to a negative absorption, i.e. the electromagnetic field which passes through the media, will be amplified rather than be dissipated.}

\added{A question that might arise is that why Re$[\sigma_{xx}]$ in the off-resonant regime is positive for all ranges of probe frequency}. \added{In the off-resonant and low amplitude of drive, because the $n$th sideband ($n\neq0$) weight is negligible with respect to zeroth sideband , these inter sideband transitions have no significant contribution to the optical conductivity and just the transitions among the lower occupied band and higher empty band in the Floquet zone have considerable effect. So, Re$[\sigma_{xx}]$ is completely
positive in this regime}; see Fig.~\ref{om1000}(b),(e). 

Indeed, as shown in Fig.~\ref{comp}(a), the transitions A, B and C are attributed to the optical transitions from the $n$th sideband to $(n+m) $th sideband, where $m=-1,0,-2$ refers to the $m$ photon processes, respectively (see Eq.~\ref{exkubo} in which there is a summation over $"m"$). It is obvious that the optical transitions between those Floquet states with low optical weights (indicated by the colored scale bar on each band in Fig.~\ref{comp}(a)) are ignorable. The reason originates from the corresponding matrix elements in Eq.~\ref{exkubo} which are too small. The peaks and dips at very low $\omega$ appeared in the components of optical conductivity, refer to the A transitions which occur in the dynamical gap.

For the sake of clarifying the relation between the transitions and optical conductivity, we decompose the summation in Eq.~\ref{exkubo} in terms of m-photon processes as represented in Fig.~\ref{comp}(b)-(e). The parameters used for this decomposition are the same as the panel (a). Note that Figs.\ref{comp} (b),(c),(d) show contribution of $m=-1,0,-2$ terms in optical conductivity. Now, it could be simply followed by eyes that the associated transitions A,B and C in Fig.~\ref{comp} (a), have their own peaks and dips in the attributed ``m''-photon processes. Moreover, the scale of each process in Fig.~\ref{comp}(b)-(e) demonstrates the importance of its contribution in optical conductivity. For example, $m=1$ processes have too small contribution compared to $m=-1$ processes. 

%%%%%%%%%%%%%%%%%%%%%%%%%%%%%%%%%%%%%%%% FIG 3 %%%%%%%%%%%%%%%%%%%%%%
\begin{figure}
\includegraphics[width= \linewidth]{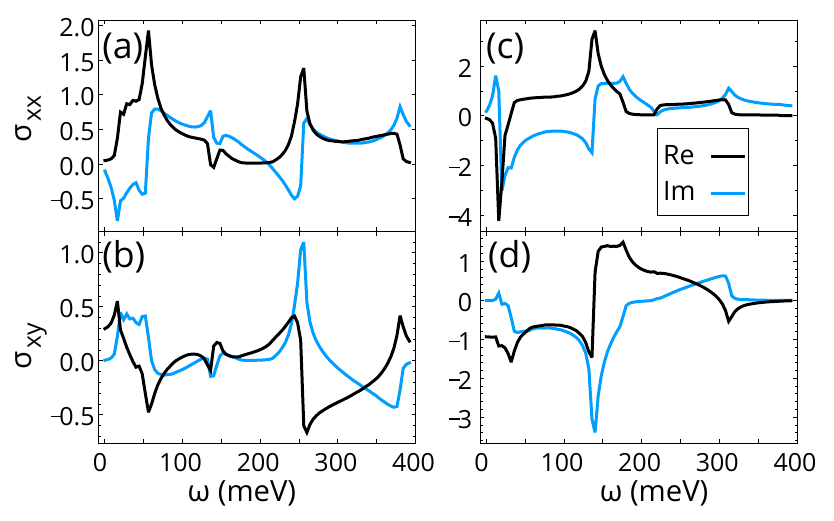} 
\caption{Time-averaged optical conductivity (in units of $e^2/h$) of 2D topological insulator irradiated by on-resonant circularly polarized light. The frequency of the drive is $\Omega=200$~meV and the intensity is $\mathcal{A}=0.10~\text{nm}^{-1}$. The \emph{mean-energy occupation} model is assumed in this calculation. The real and imaginary parts of optical conductivity are displayed with different colors. In all panels $\Delta_0=20~\text{meV}$ and $\Delta_1=55$~eV\AA$^2$ for (a)-(b) and $\Delta_1=-55$~eV\AA$^2$ for (c)-(d).}
\label{200mean}
\end{figure}

%%%%%%%%%%%%%%%%%%%%%%%%%%%%%%%%%%%%%%%%%%%%%%%%%%%%%%%%%%%%%%%%%%%%%%%%%%

%%%%%%%%%%%%%%%%%%%%%%%%%%%%%%%%%%%%%%%% FIG 3 %%%%%%%%%%%%%%%%%%%%%%

%%%%%%%%%%%%%%%%%%%%%%%%%%%%%%%%%%%%%%%%%%%%%%%%%%%%%%%%%%%%%%%%%%%%%%%%%%
\subsubsection{Mean Energy Occupation }

The simplest assumption for a non-equilibrium occupation is the mean energy distribution function, $f_{\alpha}=\theta(-\bar{\epsilon}_\alpha)$ which has been used to calculate the components of optical conductivity in Fig.~\ref{200mean} a-b (c-d) for the hopping parameters determined by the case (i) (the case (ii)). \added{As it was mentioned before, at high Fermi energies or high probe frequencies, considering the mean and quench occupation models result in a similar behavior for longitudinal optical conductivity\cite{wu2011}.}

The peaks and dips in different components of optical conductivity occur at twice of the frequency at which the van Hove singularities emerge. In fact, this correspondence fails at probe frequencies lower than band gap. The zero probe frequency limit of $Re[\sigma_{xy}(\omega \rightarrow 0]$ is not necessarily quantized anymore for this occupation, because partial occupation of the bands results in a mixture of opposite Berry curvatures giving rise to non-integer Hall conductivity. This fact is apparent from Fig.~\ref{200mean}(b),(d).
To understand these features of optical conductivity, we depict the Floquet band structure in Fig.~\ref{phmean}(a) displaying the hollow circles as the mean energy occupation and also indicating the optical weights on each band with colored scale bar. If the drive is on-resonant, in this assumption for the occupation, each band is partially filled. The reason originates from the band mixing around the momentum of the gaps. So in these k-points, the mean energy tolerates a sign change. 

The decomposition of the real part of longitudinal optical conductivity in terms of m-photon processes for the case (i) is presented in Fig.~\ref{phmean}(b)-(e). Comparison between the band structure of Fig.~\ref{phmean}(a) and corresponding optical conductivities in Fig.~\ref{phmean}(b)-(e), shows that still one can find out some optical transition for each peak or dip in the m-photon optical conductivity. 

\section{Discussion}\label{S5.5}
In this paper we assumed two simplest occupation scenarios of Floquet states, used before in literature. In experiment, the  \emph{ideal occupation} may be realized by engineering the baths \cite{seetharam} and in special conditions. Although realization of this occupation is hard in experiment, it has the advantage of showing the topology of the system through quantized Hall response. Also we  anticipate that the \emph{mean energy occupation} could be realized when the dissipation is very low, which is observed for example in engineered condensed matter systems such as cold atoms.

It should be noted that in general the occupation of Floquet states in realistic condensed matter systems is a complicated function of dissipation, heating and external parameters, which is case dependent and sensitive to special conditions of experiment. For example, in the recent experiment of light induced anomalous Hall effect in Graphene \cite{mciver} the Hall conductance of Graphene under circularly polarized light was measured. To rebuild the results in a theoretical model, \cite{nuske}, the Lindblad master equation was used with three phenomenological parameters; the decay rate, dephasing rate, and particle exchange rate with Fermionic bath. By fitting such parameters they were able to rebuild the results. Also in Ref.~\cite{seetharam} it was shown that radiative recombination and phonon-mediated relaxation, besides the Fermi and Bose baths, can be important in determination of the occupation. As a conclusion, the external effects which depend on the details of experiment are important to determine the occupation and consequently the results. 

\section{Summary}\label{S6}
In summary, we investigated the influence of a circularly polarized light on longitudinal and Hall components of time-averaged optical conductivity %in a lattice model 
of a two dimensional topological insulator. %In this calculation, 
Two simple occupation scenarios are considered:
an ideal occupation which resembles the Fermi-Dirac distribution as the static situation,
and a mean-energy occupation which simulates a phenomenological non-equilibrium situation. 
We showed that in the formaer case the irradiated light changes the topology of the system which is the origin of the quantization in the optical Hall conductivity at zero probe frequency. 
This quantization %of the optical Hall conductivity 
is absent in the mean-energy occupation case.
To interpret the peaks and dips in the optical conductivity, we follow the optical transitions at van-Hove singularities of the Floquet band structure. The intensity of each peak or dip is shown to depend strongly on the optical weight of each quasi Floquet state and also the occupation of the states. Moreover, inverted population in the Floquet states results in a negative value for the longitudinal part of optical conductivity at the on-resonant drive. 
\added{The negative longitudinal optical conductivity demonstrates an amplification of the probe light instead of the absorbance. }

\section{Acknowledgement}
We appreciate Liang Du for fruitful discussions.
\appendix 
\section{Matrix Elements of Current Operator}\label{current_op}
In the extended Kubo formula for optical conductivity Eq.~\ref{exkubo}, we have Fourier components of the current operator such as
\begin{equation}
j^{v(m)}_{\alpha\gamma}=1/T\int_0^T \langle\phi_\alpha(t)|\frac{\partial H(t)}{\partial k_v}|\phi_\gamma(t)\rangle e^{im\Omega t}dt
\end{equation}
we can expand the Floquet states $|\phi_\gamma(t)\rangle=\sum_q |\phi^q_\gamma\rangle e^{-iq\Omega t}$ and current operator $j^v(t)=\frac{\partial H(t)}{\partial k_v}=\sum_p j^{v (p)}e^{-ip\Omega t}$ in Fourier series. So, we have
\begin{equation}
\begin{aligned}
j^{v(m)}_{\alpha\gamma}=&\sum_{lpq}1/T\int_0^T \langle\phi^l_\alpha |j^{v (p)}|\phi_\gamma^q\rangle e^{i(m-q-p+l)\Omega t}dt\\
=&\sum_{lp} \langle\phi^l_\alpha|j^{v (p)}|\phi_\gamma^{l+m-p}\rangle 
\end{aligned}
\end{equation}
If the current operator is independent of time, we have
\begin{equation}
j^{v(m)}_{\alpha\gamma}=\sum_{l} \langle\phi^l_\alpha|j^{v (0)}|\phi_\gamma^{l+m}\rangle 
\end{equation}

\section{Equivalent Forms for Floquet-Kubo Formula}\label{Eqv_Kubo_Floquet}
First consider the longitudinal part of optical conductivity. Using Eq.~\ref{exkubo}, one can write
\begin{equation}
\begin{aligned}
& \sigma^{xx}(\omega)=-i \sum_{\alpha,\gamma>\alpha} \sum_{m} \frac{f_\alpha-f_\gamma}{\epsilon_\alpha-\epsilon_\gamma-m\Omega} j_{\alpha\gamma}^{x (m)}j_{\gamma\alpha}^{x (-m)} \times \\
& \big[ \frac{1}{\omega+i0^+-(\epsilon_\alpha 
-\epsilon_\gamma-m\Omega)} +\frac{ 1}{\omega+i0^++\epsilon_\alpha-\epsilon_\gamma-m\Omega}\big]
\end{aligned}
\end{equation}
Using the relation $\frac{1}{x+i 0^+}=\mathcal{P}(\frac{1}{x})-i\pi\delta(x)$ where $\mathcal{P}$ denotes the principle part, we deduce 
\begin{equation}
\begin{aligned}
&Re \sigma^{xx}(\omega)=-\pi \sum_{\alpha,\gamma>\alpha} \sum_{m} \frac{f_\alpha-f_\gamma}{\epsilon_\alpha-\epsilon_\gamma-m\Omega} |j_{\alpha\gamma}^{x (m)}|^2 \times \\
& \big[ \delta(\omega-(\epsilon_\alpha 
-\epsilon_\gamma-m\Omega)) +\delta(\omega+(\epsilon_\alpha 
-\epsilon_\gamma-m\Omega))\big]\\
& =\pi \sum_{\alpha,\gamma \neq \alpha} \sum_{m} \frac{f_\alpha-f_\gamma}{\omega} |j_{\alpha\gamma}^{x (m)}|^2 \delta(\omega+(\epsilon_\alpha 
-\epsilon_\gamma-m\Omega))
\end{aligned}
\end{equation}
This is the form used in Eq. (31) of Ref.~\cite{wu2011} where the current operator is assumed to be independent of time.

On the other hand the transversal part of Eq.~\ref{exkubo} can be written as
\begin{equation}
\begin{aligned}
& \sigma^{xy}(\omega)=-i \sum_{\alpha,\gamma>\alpha} \sum_{m} \frac{f_\alpha-f_\gamma}{\epsilon_\alpha-\epsilon_\gamma-m\Omega}\times \\
& \large\{ \frac{(\omega+i0^+)[j_{\alpha\gamma}^{y (m)}j_{\gamma\alpha}^{x (-m)}+ j_{\alpha\gamma}^{x (m)} j_{\gamma\alpha}^{y (-m)}] }{(\omega+i0^+)^2-(\epsilon_\alpha 
-\epsilon_\gamma-m\Omega)^2}\\
& + \frac{(\epsilon_\alpha 
-\epsilon_\gamma-m\Omega)[j_{\alpha\gamma}^{y (m)}j_{\gamma\alpha}^{x (-m)}- j_{\alpha\gamma}^{x (m)} j_{\gamma\alpha}^{y (-m)}] }{(\omega+i0^+)^2-(\epsilon_\alpha 
-\epsilon_\gamma-m\Omega)^2} \large\}
\end{aligned}
\end{equation} 
The first term in braces vanishes for isotropic systems (or 2D systems which have at least 3-fold rotational symmetry) because it is symmetric with respect to the $x$ and $y$ and the second term i.e. the antisymmetric term remains, so
\begin{equation}
\begin{aligned}
& \sigma^{xy}(\omega)=-i \sum_{\alpha,\gamma>\alpha} \sum_{m}(f_\alpha-f_\gamma)\times \\
& \frac{[j_{\alpha\gamma}^{y (m)}j_{\gamma\alpha}^{x (-m)}- j_{\alpha\gamma}^{x (m)} j_{\gamma\alpha}^{y (-m)}] }{(\omega+i0^+)^2-(\epsilon_\alpha 
-\epsilon_\gamma-m\Omega)^2 }
\end{aligned}
\end{equation} 
This form appeared in Eq.~(20) of Ref.~\cite{fiete2017}.


\begin{thebibliography}{99}
\bibitem{klitzing}
K. v. Klitzing, G. Dorda, and M. Pepper, Phys. Rev. Lett. 45, 494 (1980).
\bibitem{thouless}
D. J. Thouless, M. Kohmoto, M. P. Nightingale, and M. den Nijs, Phys. Rev. Lett. 49, 405 (1982).
\bibitem{Aoki2009}
T. Morimoto, Y. Hatsugai, and H. Aoki, Phys. Rev. Lett. 103, 116803 (2009).
\bibitem{haldane}
F. D. M. Haldane, Phys. Rev. Lett. 61, 2015 (1988).
\bibitem{chiu}
C. K. Chiu, J. C. Teo, A. P. Schnyder, S. Ryu, Rev. of Mod. Phys., 88(3), 035005. (2016). %Classification of topological quantum matter with symmetries. 
\bibitem{oka2009}
T. Oka and H. Aoki, Phys. Rev. B 79, 081406(R) (2009).

\bibitem{kitagawa}
T. Kitagawa, T. Oka, A. Brataas, L. Fu, E. Demler, Phys. Rev. B, 84(23), 235108 (2011). % Transport properties of nonequilibrium systems under the application of light: Photoinduced quantum Hall insulators without Landau levels.
\bibitem{opt}
G. Jotzu, M. Messer, R. Desbuquois, M. Lebrat, T. Uehlinger, D. Greif, and T. Esslinger, Nature 515, 237 (2014).
\bibitem{graphene}
J. Karch, P. Olbrich, M. Schmalzbauer, C. Zoth, C. Brinsteiner, M. Fehrenbacher, U. Wurstbauer, M. M. Glazov,
S. A. Tarasenko, E. L. Ivchenko, D. Weiss, J. Eroms, R. Yakimova, S. Lara-Avila, S. Kubatkin, and S. D. Ganichev, Phys. Rev. Lett. 105, 227402 (2010).
\bibitem{gedik}
Y. H. Wang, H. Steinberg, P. Jarillo-Herrero, and
N. Gedik, Science 342, 453 (2013).
\bibitem{dabiri1}
S. S. Dabiri, H. Cheraghchi, A. Sadeghi. Phys. Rev. B, 103, 205130 (2021).% Light-induced topological phases in thin films of magnetically doped topological insulators
\bibitem{dabiri2}
S. S. Dabiri, H. Cheraghchi, Phys. Rev. B, 104, 245121 (2021).
\bibitem{rudner}
M. S. Rudner, N. H. Lindner, E. Berg, and M. Levin, Phys. Rev. X, 3, 031005 (2013).
\bibitem{roy}
R. Roy, F. Harper, Phys. Rev. B, 96, 155118 (2017).%. Periodic table for Floquet topological insulators.
%\bibitem{prl95Q} %1 %Quantum spin hall effect in graphene.
\bibitem{FTIinvariants}
S. Yao, Z. Yan, Z. Wang. Phys. Rev. B, 96, 195303 (2017).%Topological invariants of Floquet systems: General formulation, special properties, and Floquet topological defects. 
%\bibitem{nathan}
%F. Nathan, M. S. Rudner, New J. Phys, 17(12), 125014,(2015). % Topological singularities and the general classification of Floquet–Bloch systems.
\bibitem{rudner_review}
M. S Rudner, N. H. Lindner, Nat. rev. phys., 2, 229-244(2020).



\bibitem{foa2014}
L. E. F. F. Torres, P. M. Perez-Piskunow, C. A. Balseiro,8 and G. Usaj, Phys. Rev. Lett. 113, 266801 (2014).
\bibitem{BHZ}
B. A. Bernevig, T. L. Hughes, and S.-C. Zhang, Science 314, 1757 (2006).
\bibitem{coldBHZ}
Q.X. Lv, Y.X. Du, Z.T. Liang, H.Z. Liu, J.H. Liang, L.Q. Chen, L.M. Zhou, S.C. Zhang, D.W. Zhang, B.Q. Ai, and H. Yan, Phys. Rev. Let., 127 , 136802 (2021). %Measurement of spin Chern numbers in quantum simulated topological insulators.
\bibitem{universality}%29
J. Wang, B. Lian and S. C. Zhang, Phys. Rev. B. {\bf 89}, 085106 (2014).
\bibitem{effective} %20 %effective Hamiltonian
W. Y. Shan, H. Z. Lu and S. Q. Shen, New J. Phys {\bf 12}, 043048 (2010).
\bibitem{geometrical}
D. Sticlet, F. Piechon, J. N. Fuchs, P. Kalugin, and P. Simon, Phys. Rev. B. {\bf 85}, 165456 (2012).
\bibitem{PRL2013_zhang}%20
J. Wang, B. Lian, H. Zhang, Y. Xu, and S. C. Zhang, Phys. Rev. Lett. {\bf 111}, 136801 (2013) .
\bibitem{shirley}
J. H. Shirley, Phys. Rev. 138, B979 (1965).
\bibitem{wu2011}
Y. Zhou, M. W. Wu, . Phys. Rev. B, 83 , 245436 (2011).%Optical response of graphene under intense terahertz fields
\bibitem{Discrete_BZ}
T. Fukui, Y. Hatsugai, H. Suzuki, J. Phys. Soc. Japan, {\bf 74} 1674 (2005).
\bibitem{Bukov}
M. Bukov, L. D'Alessio, A. Polkovnikov, Advances in Physics, 64(2), 139-226 (2015). %Universal high-frequency behavior of periodically driven systems: from dynamical stabilization to Floquet engineering.
\bibitem{eckart}
A. Eckardt, E. Anisimovas, New j. Phys. , 17(9), 093039 (2015).% High-frequency approximation for periodically driven quantum systems from a Floquet-space perspective.
\bibitem{bw}
T. Mikami, S. Kitamura, K. Yasuda, N. Tsuji, T. Oka, H. Aoki, Phys. Rev. B, 93(14), 144307 (2016).% Brillouin-Wigner theory for high-frequency expansion in periodically driven systems: Application to Floquet topological insulators.

\bibitem{oka2010}
T. Oka and H. Aoki, J. Phys.: Conf. Ser. 200, 062017 (2010).
\bibitem{dehghaniOC}
 H. Dehghani, A.  Mitra, Phys. Rev. B, 92, 165111 (2015). %Optical Hall conductivity of a Floquet topological insulator.
\bibitem{seetharam}
K. I. Seetharam, C-E Bardyn, N. H. Lindner, M. S. Rudner, and G. Refael, Phys. Rev. X, 5, 041050 (2015).
\bibitem{seradjeh2020}
Kumar, A., Rodriguez-Vega, M., Pereg-Barnea, T., Seradjeh, B. Phys. Rev. B, 101, 174314 (2020). .%Linear response theory and optical conductivity of Floquet topological insulators.
\bibitem{mciver}
J. W. McIver, B. Schulte,  F. U. Stein, T. Matsuyama, G. Jotzu, G. Meier, A. Cavalleri,   Nat. phys., 16, 38-41 (2020). %Light-induced anomalous Hall effect in graphene.
\bibitem{nuske}
 M. Nuske, L. Broers,  B. Schulte,  G. Jotzu,  S. A. Sato, A. Cavalleri, A Rubio, J. W.  McIver, and L. Mathey (2020). %Floquet dynamics in light-driven solids. Physical Review Research, 2(4), 043408.


\bibitem{fiete2017}
L. Du, X. Zhou, G. A. Fiete, Phys. Rev. B , 95, 035136 (2017).

\iffalse



%C. L. Kane and E. J. Mele, Phys. Rev. Lett. {\bf 95}, 226801 (2005).
%\bibitem{prl96} %2 %Quantum spin hall effect.
%B. A. Bernevig and S. C. Zhang, Phys. Rev. Lett. {\bf 96}, 106802 (2006).

%\bibitem{prl97} %4 %Quantum spin hall effect and enhanced magnetic response by spin–orbit coupling.
%S. Murakami, Phys. Rev. Lett. {\bf 97}, 236805 (2006).
%\bibitem{science314} %5 % Quantum spin hall effect and topological phase transition in HgTe 2D TIs.
%B. A. Bernevig, T. L. Hughes and S. C. Zhang, Science {\bf 314}, 1757–1761 (2006).
%\bibitem{science318} %6 %Quantum spin hall insulator state in HgTe 2D TIs.
%M. König, S. Wiedmann, C. Brüne, A. Roth, H. Buhmann, L. W. Molenkamp, X. L. Qi, S. C. Zhang, Science {\bf 318}, 766–770 (2007).
%\bibitem{nature398} %11
%Y. Xia, D. Qian, D. Hsieh, L. Wray, A. Pal, H. Lin, A. Bansil, D. Grauer, Y. S. Hor, R. J. Cava, M. Z. Hasan, Nat. Phys. {\bf 5}, 398 (2009).
%\bibitem{science329} %5 %QAH TI
%R. Yu, Wei Zhang, H. J. Zhang, S. C. Zhang, X. Dai, Z. Fang, Science {\bf 329}, 61 (2010).
%\bibitem{science340} %6 %QAH TI
%C. Z. Chang, \textit{et. al.}, Science {\bf 340}, 167 (2013).
%\bibitem{prl113137201}%16 %QAH TI
%X. Kou, \textit{et. al.}, Phys. Rev. Lett. {\bf 113}, 137201 (2014).
%\bibitem{nature10731} %17 %QAH TI
%J. G. Checkelsky, \textit{et. al.}, Nat. Phys. {\bf 10}, 731 (2014).



%\bibitem{prb81H} %18 %Massive Dirac fermions and spin physics in an ultrathin film of topological insulator.
%H. Z. Lu, W. Y. Shan, W. Yao, Q. Niu, and S. Q. Shen, Phys. Rev. B. {\bf 81}, 115407 (2010).
%\bibitem{prl2013QT} %19
%H. Z. Lu, A. Zhao and S. Q. Shen, Phys. Rev. Lett. {\bf 111}, 146802 (2013).
%\bibitem{zeemanelec}
%H. Li, L. Sheng, D. Y. Xing, Phys. Rev. B, 85(4), 045118 (2012). %Quantum phase transitions in ultrathin films of three-dimensional topological insulators in the presence of an electrostatic potential and a Zeeman field. .
%\bibitem{zyuzin}
%A. A. Zyuzin, A. A. Burkov, Phys. Rev. B, 83(19), 195413 (2011). %Thin topological insulator film in a perpendicular magnetic field. 
%\bibitem{edge}
%S. B. Zhang, H. Z. Lu, S. Q. Shen, Sci. rep., 5, 13277 (2015). %Edge states and integer quantum Hall effect in topological insulator thin films. 
\bibitem{feng}
S. F. Zhang, H. Jiang, X. C. Xie, Q. F. Sun, Phys. Rev. B. {\bf 89}, 155419 (2014). 
\bibitem{QHETI}
H. Li, L. Sheng, D. Y. Xing, Phys. Rev. B, 84(3), 035310 (2011). %Quantum Hall effect in thin films of three-dimensional topological insulators.
\bibitem{2013Gomez}
A. Gomez-Leon, and G. Platero, Phys. Rev. Lett. 110, 200403 (2013).% "Floquet-Bloch theory and topology in periodically driven lattices."
\bibitem{exp_haldane}
G. Jotzu, M. Messer, R. Desbuquois, M. Lebrat, T. Uehlinger,D. Greif, and T. Esslinger, Nature, {\bf 515}, 237 (2014). %“Experimental realisation of the topo-logical haldane model,”
\bibitem{2013Wang}
Y. H. Wang, \textit{et, al.}, Science 342.6157 453-457 (2013). %"Observation of Floquet-Bloch states on the surface of a topological insulator." 
\bibitem{2}
M. D. Reichl, and E. J. Mueller, Phy. Rev. A. 89(6) 063628 (2014). % Floquet edge states with ultracold atoms.
\bibitem{2011Gu}
Z. Gu, \textit{et, al.}, Phys. Rev. Lett.107, 216601 (2011). %"Floquet spectrum and transport through an irradiated graphene ribbon."
\bibitem{2014Perez}
P. M. Perez-Piskunow, \textit{et, al.} Phys. Rev. B 89, 121401(R) (2014).% "Floquet chiral edge states in graphene." 
\bibitem{2013Delplace}
P. Delplace,\textit{et, al.}, Phys. Rev. B 88 245422 (2013). %"Merging of Dirac points and Floquet topological transitions in ac-driven graphene."
\bibitem{2013Cayssol}
J. Cayssol, B. Dora, F. Simon, and R. Moessner, Phys.Status Solidi RRL 7, 101 (2013).
\bibitem{2011Lindner}
N. H. Lindner, \textit{et, al.}, Nature Physics 7.6 490 (2011). %"Floquet topological insulator in semiconductor 2D TIs." 
\bibitem{2010Kitagawa}
T. Kitagawa, \textit{et, al.}, Phys. Rev. B 82, 235114 (2010).% "Topological characterization of periodically driven quantum systems." 
\bibitem{2016Lovey}
D. A. Lovey, \textit{et, al.}, Phys. Rev. B 93, 245434 (2016). %"Floquet bound states around defects and adatoms in graphene."

\bibitem{creat_edge}
L. Zhou and J. Gong, Phys. Rev. A 97, 063603 (2018).

\bibitem{nature584}
Zhang, Yi, et al, Nat. Phys. {\bf 6}, 584 (2010).
\bibitem{prb81041307} %14 %finitegap %Oscillatory crossover from two dimensional to three dimensional topological insulators.
C. X. Liu, H. J. Zhang, B. Yan, X. L. Qi, T. Frauenheim, X. Dai, Z. Fang, S. C. Zhang, Phys. Rev. B. {\bf 81}, 041307(R) (2010).
\bibitem{Magnetic}%28
V. Kulbachinskii, P. M. Tarasov. E. Brük, JETP Letters, {\bf 73}, 352 (2001).




\bibitem{Prodan}%32
E. Prodan, Phys. Rev. B 80, 125327 (2009).
\bibitem{Sheng}%32
D. N. Sheng, Z. Y. Weng, L. Sheng, and F. D. M. Haldane,
Phys. Rev. Lett. 97, 036808 (2006).
\bibitem{spinchernnumberTI}
H. Li, L. Sheng, D. N. Sheng, D. Y. Xing, Phys. Rev. B 82 165104 (2010).
\bibitem{spinchernnumber}
Sheng Li et al, Chinese Phys. B 22 067201 (2013).
\bibitem{prl95Z} %3 %Z2 topological order and the quantum spin hall effect.
C. L. Kane and E. J. Mele, Phys. Rev. Lett. {\bf 95}, 146802 (2005).
%\bibitem{Optical_NLSM1}
%S. P. Mukherjee and J. P. Carbotte, Phys. Rev. B. {\bf 95}, 214203 (2017).
%\bibitem{Optical_NLSM2}
%S. Barati, S. H. Abedinpour, Phys. Rev. B. {\bf 96}, 155150 (2017).
%\bibitem{Magnetic_NLSM}
%H-J. Duan, S. H. Zheng, Y. Y. Yang, C. Y. Zhu, M. X. Deng, M. Yang, R. Q. Wang, Phys. Rev. B. {\bf 102}, %165110 (2020).
\bibitem{kubofinite}
P. Dutta, S. K. Maiti, S. N. Karmakar, J. Appl. Phys., {\bf 112}(4), 044306 (2012).
\bibitem{chen2012}
J. C. Chen, J. Wang, Q. F. Sun, Phys. Rev. B, 85(12), 125401 (2012).%Effect of magnetic field on electron transport in HgTe/CdTe 2D TIs: Numerical analysis. Physical Review B, 85(12), 125401.

\fi




\end{thebibliography}
\end{document}